\documentclass[journal]{IEEEtran}
\ifCLASSINFOpdf
\else
\fi
\usepackage{tabularx}
\usepackage{orcidlink}
\usepackage{float}
\usepackage{cite}
\usepackage{amsmath,amssymb,amsfonts}
\usepackage{algorithmic}
\usepackage{graphicx}
\usepackage{graphicx}
\usepackage{textcomp}
\usepackage{xcolor}
\usepackage{hyperref}
\usepackage{makecell}
\usepackage{multirow}
\usepackage{threeparttable}
\usepackage[ruled,vlined, linesnumbered]{algorithm2e}
\newcommand{\etal}{\textit{et al.}}

\begin{document}
%
\title{Efficient Memristive Spiking Neural Networks Architecture with Supervised In-Situ STDP Method
}


\author{%
Santlal Prajapati \orcidlink{0000-0002-8588-812X},
Susmita Sur-Kolay \orcidlink{0000-0002-2052-3779},~\IEEEmembership{Senior Member,~IEEE,}
and Soumyadeep Dutta \orcidlink{0000-0002-1682-4990}%
\thanks{Santlal Prajapati is with the Advanced Computing and Microelectronics Unit, Indian Statistical Institute, Kolkata 700108, India.
(e-mail: santlal.santy@gmail.com).}
\thanks{Susmita Sur-Kolay is a retired Professor with the Advanced Computing and Microelectronics Unit, Indian Statistical Institute, Kolkata 700108, India (e-mail: surkolay@gmail.com).}
\thanks{Soumyadeep Dutta is with Balasore Alloys Limited, Balgopalpur, Odisha 756020, India (e-mail: soumyadeepd@edpi.ai).}
\thanks{This work was carried out at the Advanced Computing and Microelectronics Unit, Indian Statistical Institute, Kolkata 700108, India.}

\thanks{© 2026 IEEE. Personal use of this material is permitted. Permission from IEEE must be obtained for all other uses, in any current or future media, including reprinting/republishing this material for advertising or promotional purposes, creating new collective works, for resale or redistribution to servers or lists, or reuse of any copyrighted component of this work in other works.}

}


%



\IEEEtitleabstractindextext{%
\begin{abstract}

Memristor-based Spiking Neural Networks (SNNs) with temporal spike encoding enable ultra-low-energy computation, making them ideal for battery-powered intelligent devices. This paper presents a circuit-level memristive spiking neural network (SNN) architecture trained using a proposed novel supervised in-situ learning algorithm inspired by spike-timing-dependent plasticity (STDP). The proposed architecture efficiently implements lateral inhibition and the refractory period, eliminating the need for external microcontrollers or ancillary control hardware. All synapses of the winning neurons are updated in parallel, enhancing training efficiency. The modular design ensures scalability with respect to input data dimensions and output class count. The SNN is evaluated in LTspice for pattern recognition (using 5×3 binary images) and classification tasks using the Iris and Breast Cancer Wisconsin (BCW) datasets. During testing, the system achieved perfect pattern recognition and high classification accuracies of 99.11\% (Iris) and 97.9\% (BCW). Additionally, it has demonstrated robustness, maintaining an average recognition rate of 93.4\% under 20\% input noise. The impact of cycle-to-cycle variation, line resistance, parasitic capacitance, stuck-at-conductance faults, and memristor device variations was also analyzed. In addition, LTspice simulations demonstrated the unsupervised behavior of the proposed memristive SNN, where post-synaptic neurons autonomously associated themselves with different classes of the IRIS dataset.


\end{abstract}

}

\maketitle

\IEEEdisplaynontitleabstractindextext

%
\IEEEpeerreviewmaketitle

\section{Introduction}
\label{intro}

The primary focus of contemporary software systems is on Artificial Intelligence (AI) tasks \cite{abiodun2018state}. A fast and power-efficient hardware platform is essential for implementing Artificial Neural Networks (ANNs), the primary technologies that enable AI tasks. Adopting biologically plausible information processing primitives, such as spiking neural networks (SNN), is recommended as a power-efficient platform \cite{roy2019towards}. SNNs are more power-efficient than ANNs due to their spike-driven sparse computation feature. These third-generation neural networks~\cite{maass1997networks} exhibit event-driven computation, diverse coding mechanisms, and rich neurodynamic characteristics in both space and time~\cite{roy2019towards}.

\begin{table*}[t]
\caption{Qualitative comparison of the proposed architecture with related prior works}
\label{tab:comparison}
\centering
\resizebox{1\linewidth}{!}{%
\begin{tabular}{|c| c| c| c| c| c|c |c| c|}
\hline
\textbf{Papers} &
\textbf{Synapse} &
\begin{tabular}[c]{@{}c@{}}\textbf{Learning}\\ \textbf{algorithm}\end{tabular} &
\begin{tabular}[c]{@{}c@{}}\textbf{The way to update}\\ \textbf{the memristive}\\ \textbf{synapses of MC}\end{tabular} &
\textbf{WTA} &
\begin{tabular}[c]{@{}c@{}}\textbf{Refractory}\\ \textbf{period}\end{tabular} &
\begin{tabular}[c]{@{}c@{}}\textbf{Image recognition and}\\ \textbf{\# training Samples}\\ \textbf{to get same accuracy}\end{tabular} &
\begin{tabular}[c]{@{}c@{}}\textbf{Major hardware components}\\ \textbf{affecting power and area}\end{tabular} &
\textbf{Problems solved} \\
\hline

\cite{zhao2018neurocomputing} &
2 Op, 4T, 1M &
STDP Unsupervised & Parallel &
No & No &
No, ------ &
\begin{tabular}[c]{@{}l@{}}Two Op per synapse\end{tabular} &
Pattern recognition \\

\hline

\cite{zhang2021ScienceBulletin} &
1T1M &
\begin{tabular}[c]{@{}l@{}}Hebbian Unsupervised at\\ first layer and supervised\\ at second layer\end{tabular} &
Parallel &
Yes & Yes &
Ten 6x5 digits, ----- &
\begin{tabular}[c]{@{}l@{}}Flip-flops, AND gates,\\ and buffer in neurons\end{tabular} &
Pattern recognition \\

\hline

\cite{li2021ICIST} &
1M &
Unsupervised STDP &
Parallel &
Yes & No &
Four 5x3 digits, 3000 &
\begin{tabular}[c]{@{}l@{}}$(m-1)$ NMOS T, two OP\\ and Two 74121 ICs\\ (monostable multivibrators)\\ per post-neuron\end{tabular} &
Pattern recognition \\

\hline

\cite{li2022nca} &
\begin{tabular}[c]{@{}c@{}}1M 1 Dual Switch\end{tabular} &
\begin{tabular}[c]{@{}l@{}}Hebbian Unsupervised\end{tabular} &
Parallel &
Yes & No &
Four 5x3 letters, 400 &
\begin{tabular}[c]{@{}l@{}}One OP, $(m-1)$ PMOS T\\ per post-neuron\end{tabular} &
Pattern recognition \\

\hline

\cite{sboev2021mathematics} &
------ &
Supervised STDP &
Parallel (Theoretically) &
Yes & Yes &
No, ----- &
Mathematical model &
Classification \\

\hline

\cite{jiang2022memristor} &
2M &
SpikeProp~\cite{bohte2002error} &
Parallel &
No & No &
No, ---- &
\begin{tabular}[c]{@{}l@{}}555-timer and three C per neuron.\\ Delay circuit per synapse\\ multiple synapses between\\ pre- and post-neurons\end{tabular} &
Classification \\

\hline

\cite{Liu_2024_tvlsi} &
4M2R &
Unsupervised STDP &
Parallel &
No & No &
$3 \times 3$ patterns &
\begin{tabular}[c]{@{}l@{}}Synapses and subtractors\\ STDP circuits and trace modules\end{tabular} &
pattern recognitions \\

\hline

\cite{Verma2025tcas} &
\begin{tabular}[c]{@{}c@{}}fully complex\\ CMOS synapse\end{tabular} &
\begin{tabular}[c]{@{}l@{}}On-chip STDP\\ (unsupervised)\end{tabular} &
Parallel &
No & Yes &
XOR and MNIST &
\begin{tabular}[c]{@{}l@{}}CMOS-based\\ memristor emulators,\\ Spike Processing Circuitry\end{tabular} &
classification \\

\hline

\cite{Xin2024} &
\begin{tabular}[c]{@{}c@{}}1M\end{tabular} &
\begin{tabular}[c]{@{}l@{}}unsupervised STDP\end{tabular} &
Parallel &
No & Yes &
$3 \times 3$ pattern &
\begin{tabular}[c]{@{}l@{}}DAC/ADC STDP timing circuits,\\ neuron circuits, 555 timer \end{tabular} &
pattern recognitions \\

\hline

\cite{Ma2023} &
\begin{tabular}[c]{@{}c@{}}2T2M\end{tabular} &
\begin{tabular}[c]{@{}l@{}}unsupervised STDP\end{tabular} &
Parallel &
No & Yes &
$8 \times 8$ pattern &
\begin{tabular}[c]{@{}l@{}}DAC/ADC, circuit \\controller, Hidden layer\end{tabular} &
pattern recognitions \\

\hline

\textbf{This work} &
\textbf{1T1M} &
\textbf{Supervised STDP} &
\textbf{Parallel} &
\textbf{Yes} & \textbf{Yes} &\textbf{
\begin{tabular}[c]{@{}c@{}}four 5x3 digits, 100\\ 7x3 digits\end{tabular}} &\textbf{
\begin{tabular}[c]{@{}l@{}}Three OPs per post-neuron\\ One comparator per pre-neuron\end{tabular}} &\textbf{
\begin{tabular}[c]{@{}c@{}}Pattern recognition\\ and Classification\end{tabular}} \\

\hline
\end{tabular}
}
\footnotesize
\begin{tablenotes}
       \item  Op: Operational amplifier, T: Transistor, M: Memristor, R: Resistor, C: Capacitor, WTA: Winner-Takes-All, $m$:Number of post-neurons, MC: Memristive Crossbar
     \end{tablenotes}

\end{table*}

In order to achieve brain-inspired efficiency, SNNs require a hardware architecture that supports synaptic plasticity, high parallelism, high-density integration, and low-power devices. The nano-sized, nonlinear memristor devices, first fabricated in HP Labs in 2008~\cite{strukov2008missing}, are promising candidates for SNN hardware design. Memristors~\cite{chuaMem} are passive, nonlinear circuit elements with variable nonvolatile resistance states, similar to biological synapses. They are compatible with CMOS technology~\cite{jiang2019integrating}. In SNN circuit implementations, memristors enable the co-location of computation and memory~\cite{mutlu2022modern}, thus alleviating the memory bottleneck problem~\cite{wulf_1995_ACM_hitting}, a major impediment to data-intensive ANN workloads.

Existing research~\cite{pedretti2017SR, hansen2018Scientificreports, Jiang2018ITCAS, li2021ICIST, li2022nca, zhao2018neurocomputing, zhang2021ScienceBulletin, hajiabadi2021memristor,Wang_ITBCS_2023,Liu_2024_tvlsi,Verma2025tcas,zhou_iscas_2022,Wijesinghe2018tetci, Ma2023,Xin2024} has explored in-situ trainable circuit implementations of memristive SNNs for high processing speed, low power dissipation, and high parallelism. In-situ training is performed on neuromorphic hardware itself~\cite{alibart_2013_nature_communication}. However, related works~\cite{pedretti2017SR, hansen2018Scientificreports, Jiang2018ITCAS,Wang_ITBCS_2023} require additional control units like micro-controllers or FPGAs. Although some memristive SNNs with in-situ training mechanisms eliminate external control units, their hardware complexity remains high~\cite{li2021ICIST, li2022nca, zhao2018neurocomputing, zhang2021ScienceBulletin}. In a memristive SNN architecture, the simplicity of the artificial synapse circuit is crucial for scalability. In~\cite{zhao2018neurocomputing}, the artificial synapse circuit using commercial ICs and discrete semiconductor components is complex. Similarly, the synapse update circuit in~\cite{li2021ICIST} uses commercial ICs, making the solution less economical. 

The winner-takes-all (WTA) mechanism, where only one neuron spikes while the others are laterally inhibited, is attractive for low-power neuromorphic computation applications. While WTA has been implemented~\cite{zhang2021ScienceBulletin,zhou_iscas_2022} using a separate memristive crossbar array and mixed-signal peripheral circuits~\cite{zhang2021ScienceBulletin}, we propose a complete analog approach without any extra memristive crossbar for this strategy. Furthermore, the WTA architectures proposed in~\cite{li2021ICIST, li2022nca} have post-synaptic neurons connected in a mesh topology, hindering scalability. We propose a more scalable and hardware-efficient alternative.

In ~\cite{li2021ICIST, li2022nca, zhao2018neurocomputing, zhang2021ScienceBulletin, Ma2023,Xin2024}, researchers have focused on pattern recognition tasks, while others~\cite{sboev2021mathematics, jiang2022memristor} have addressed classification problems using memristor-based SNNs. For instance, Sboev et al.~\cite{sboev2021mathematics} proposed a mathematical model of memristive SNNs for classification tasks, whereas Jiang et al.~\cite{jiang2022memristor} suggested a hardware architecture of memristive SNNs with a hidden layer, implementing WTA and refractory period as computational primitives. The hardware complexity of Jiang's memristive SNN architecture~\cite{jiang2022memristor} is a concern due to its multi-synaptic model and the use of commercial ICs. The authors of ~\cite{Xin2024} proposed a trainable memristive SNN using transistor-less synapses, making the crossbar susceptible to sneak-path currents, while also employing DAC/ADC interfaces and 555-timer circuits that increase hardware complexity. Similarly, in ~\cite{Ma2023} 2T2M (two transistors, two memristors) synapses, hidden layers, and DAC/ADC interfaces, result in higher hardware complexity compared to the proposed compact fully analog 1T1M architecture.

Our memristive SNN architecture is simpler at the circuit-level, has no hidden layer, and uses 1T1M  synapse.

Our architecture can be used for pattern recognition as well as classification problems. Identification of a structure in an image input is termed as pattern recognition~\cite{li2021ICIST, li2022nca, zhao2018neurocomputing, zhang2021ScienceBulletin}, whereas assigning an input sample, which may not be an image, to one of the several predefined categories or classes~\cite{sboev2021mathematics, jiang2022memristor,Verma2025tcas} is called a classification problem. For classification, inputs are preprocessed with the Gaussian receptive field method~\cite{sboev2021mathematics}, which increases robustness against faulty memristors. Inputs are encoded into spikes using the temporal spike encoding method~\cite{sboev2021mathematics}. Temporal encoding and sparse synapse updating (only the winner's synapses are updated in parallel) make the proposed SNN more time-energy-efficient compared with one having rate encoding of spikes \cite{rateencoding_diehl2015unsupervised}.

Table~\ref{tab:comparison} provides a qualitative comparison with similar previous works solving pattern recognition and classification problems, highlighting the advantages of the proposed approach in terms of hardware complexity and statistical efficiency.

The complete and simple, hardware and training latency efficient, in-situ trainable, and external microcontroller-free memristive SNN architecture that solves pattern recognition and classification problems are the motivations of this work. 
In this work, we contribute the following:
\begin{itemize}
    \item a complete circuit-level architecture of a trainable and scalable memristive SNN that does not require any external control units such as a microcontroller or an FPGA,

    \item a supervised in-situ STDP training algorithm is proposed,
    
    \item design of peripheral/control circuits to accomplish training, emulate WTA strategy and refractory period while achieving high accuracy and F1-score,
    
        \item thorough study of the effects of faulty memristors and device parameter variations, including $R_{on}$, $R_{off}$, and memristor threshold voltage, on the performance of the proposed SNN. In addition, the effects of practical non-idealities such as cycle-to-cycle variability, line resistance, and parasitic capacitance on the proposed architecture are also analyzed.
    
    \item an analysis of robustness against noisy patterns in the memristive SNN.
\end{itemize}

The remainder of this article is organized as follows. Section~\ref{prilims} presents the SNN model, memristor, memristive crossbar, and data pre-processing. Section~\ref{archi} describes the proposed memristive SNN architecture and its components. Section~\ref{train} discusses the supervised in-situ STDP training algorithm, while Section~\ref{simulation} presents simulation results. Finally, Section~\ref{conclusion} concludes the article.

\section{Preliminaries}
\label{prilims}
\subsection{Spiking Neural Networks}
In a spiking neuronal network, a spike or an action potential is a short electrical pulse used for inter-neuron communication~\cite{wulfram_2002_cambridge}. Each spiking neuron has a membrane potential that increases upon receiving a spike and gradually decreases afterward. When there are no input spikes, then it stays at resting potential. A neuron fires a spike when its membrane potential crosses the threshold potential, and afterward, it goes below resting potential~\cite{wulfram_2002_cambridge}. Once a neuron fires a spike, it waits for some time, called the refractory period, to fire again. In a layer of neurons, when a neuron fires, it inhibits others from firing. This is called lateral inhibition, which makes the spike communication sparse.

\begin{table}[]
\caption{The parameter values used in the spice memristor model.}
\label{tab:memristor_parameter}
  \resizebox{\linewidth}{!}{%
\begin{tabular}{|l|l|l|l|l|l|l|l|l|l|}
\hline 
\begin{tabular}[c]{@{}l@{}}Spice model \\ parameters\end{tabular} &
  D(nm) &
  $\mu_{v}(m^{2}s^{-1}\Omega^{-1})$ &
  $R_{on}(\Omega$) &
  $R_{off}(\Omega$) &
  $V_{T-}(V)$ &
  $V_{T+}$(V) &
  $i_{on}(A)$ &
  $i_{off}(A)$ &
  $i_{0}(A)$ \\ \hline
Values &
  $3$ &
  $3.2e-15$ &
  $1e6$ &
  $6e7$ &
  $-2.4$ &
  $1.2$ &
  $1$ &
  $1.4e-14$ &
  $3e-8$ \\ \hline
\end{tabular}}
\end{table}

\begin{figure}[h]
    \centering
    \includegraphics[width=0.8\linewidth]{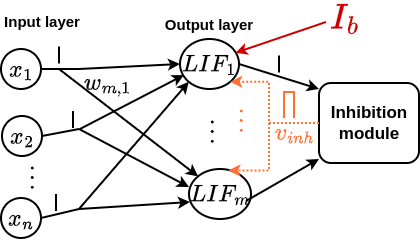}
    \caption{Spiking neural network (SNN) model. Input features are encoded into spikes (vertical lines) and propagated through synapses to post-synaptic neurons in the output layer. The $m^{\text{th}}$ neuron is denoted as $LIF_m$. The inhibition module implements winner-take-all (WTA) and refractory behavior via the inhibition signal $v_{inh}$.}
    \label{fig:snn}
\end{figure}

The spiking neural network (SNN) model employed in this work is shown in Fig.~\ref{fig:snn}. It is simple and has no hidden layers, performing both classification and pattern recognition tasks. The working process of it is as follows ---
\begin{itemize}
    \item The input feature vector $\mathbf{x}=[x_{1},~x_{2},~\hdots,~x_{n}]$, at the input layer, is encoded into spikes, specifically called pre-synaptic spikes or here $pre\_spikes$, using the temporal spike encoding method explained in Section~\ref{temp}.
    
    \item These $pre\_spikes$ travel via weighted synapses to neurons at the output layer called post-synaptic neurons or here $post\_neuron$s that are denoted as $LIF_{1}$ to $LIF_{m}$. The synapse $s_{m,1}$ denotes the connection between $post\_neuron_{m}$ and $pre\_neuron_{1}$ (or $pre\_spike_{1}$) with synaptic weight $w_{m,1}$.
    
    \item The post-neurons are competitive; they compete amongst themselves to fire first and laterally inhibit others using the inhibition module.

    \item The bias current $I_{b}$ is applied selectively to a particular neuron, based on the input sample level, thereby enabling supervised training.
   
   \item The neuron that spikes first sends its $post\_spike$ to the inhibition module, and in response to that, the inhibition module sends an inhibition pulse signal $v_{inh}$ to all LIFs to prevent their firings.
   
   \item  The neuron that spikes first is called the winner, and this process is called winner-takes-all (WTA). The $v_{inh}$ plays a vital role in implementing the refractory period, discussed in Section~\ref{LIC}.
\end{itemize}
For a dataset, if there are $n$ features and $m$ classes, then there are $n$ $pre\_spikes$, $m$ $post\_neurons$, and $n \times m$ weighted synaptic connections. The SNN models are trained with the proposed supervised algorithms based on STDP~\cite{linares2009memristance} (detailed in Section \ref{train}).

\subsection{Memristor model and memristive crossbar}
\label{memristor} 
Memristors, due to their nano-scale size and variable non-volatile conductance states, are the most suitable devices to implement artificial synapses and emulate synaptic plasticity in neuromorphic systems. The spice model of voltage-controlled threshold memristor proposed by Yang \etal \cite{Zhang_tcas_2017} is chosen to design a synapse in this work. The mathematical expressions of memristors are:
\begin{eqnarray}
    \label{basic}
    \begin{cases}
        v(t)=R(t)i(t)\\
    R(t)=R_{on}\frac{w(t)}{D}+R_{off}(1-\frac{w(t)}{D}) 
    \end{cases}
\end{eqnarray}
where, $v(t),~i(t),~R(t),~D,~R_{on},~R_{off},~and~ w(t)$ are voltage, current, memristance of memristor, width of memristive device, minimum memristance, maximum memristance, and width of doped region in the memristive device respectively. The dynamics of $w(t)$ is as follows:
\begin{equation}
\label{dynamic}
\frac{dw(t)}{dt}=
    \begin{cases}
        \mu_{v} \frac{R_{on}}{D} \frac{i_{off}}{i(t)-i_{0}} f(w(t)), & v(t)>V_{T+}>0 \\
        0, & V_{T-} \le v(t) \le V_{T+} \\
        \mu_{v} \frac{R_{on}}{D} \frac{i_{(t)}}{i_{on}} f(w(t)),& v(t)<V_{T-}<0
    \end{cases}
\end{equation}
where, $\mu_{v},~ f(w(t)),~V_{T+},~ and ~V_{T-}$ are average ion mobility, window function, and positive \& negative threshold voltages respectively. The $~i_{on},~i_{off},~and ~ i_{0}$ are fitting parameters. The window function $f(\cdot)$ is given by:
\begin{equation}
    \label{window}
    f(w(t))=1-{\left(\frac{2w(t)}{D}-1\right)}^{2p}
\end{equation}
Where $p$ is a positive constant. The parameters' values used in this work are listed in \autoref{tab:memristor_parameter}, which matches the characteristics of $PCMO$-based practical memristor device~\cite{SHERI_2014_itie,chu2014ITIE}. A positive voltage across the memristor, greater than the positive threshold, increases its conductance and vice versa ~\cite{Zhang_tcas_2017}. 

The structure where memristors are organized in a 2D array is called memristive crossbar (MC)~\cite{yakopcic_IJCNN_2011}. On MC, memristors are placed alone (1M type, 2M type, 2M1M types)~\cite{li2021ICIST,hasan2019ex,Teimoori_2018tvlsi} or with transistors~\cite{yakopcic_IJCNN_2011,shahar15} (1T1M or 2T1M types, etc). The 1T1M type crossbar is shown in the green dashed box of Fig.~\ref{fig:overview_architecture}(a). It is used to perform matrix-vector multiplication when MC is considered a matrix of memristors' conductance and the vector is voltages to the rows of MC~\cite{kim2024memristor,hu2016dot}.

\subsection{Data Pre-processing}
\label{preprocessing}
For classification problems, the input features are pre-processed by min-max scale followed by Gaussian receptive field methods~\cite{sboev2020solving}. The Gaussian receptive field method expands an input vector $\mathbf{x}$ (originally $n_1$-dimensional) to a higher-dimensional space ($n_{1}.n_{2}$) where $n_{2}$ is number of receptive fields. Each element $x_{i}$ of $\mathbf{x}$ is expanded to $n_{2}$ components~\cite{sboev2021mathematics, sboev2020solving}. Only for classification simulations, we perform Gaussian receptive field pre-processing with $n_{2}=3$, but not for pattern recognition. After pre-processing, it is assumed that the minimum and maximum values of each feature are $0$ and $T$, respectively. This method has been widely used in SNN-based models~\cite{sboev2020solving, sboev2021mathematics, wang2014neurocomputing, yu2014Neurocomputing, gutig2006tempotron}.

\section{Memristive SNN Architecture}
\label{archi}

\begin{figure}[h]
    \centering
    \includegraphics[width=\linewidth]{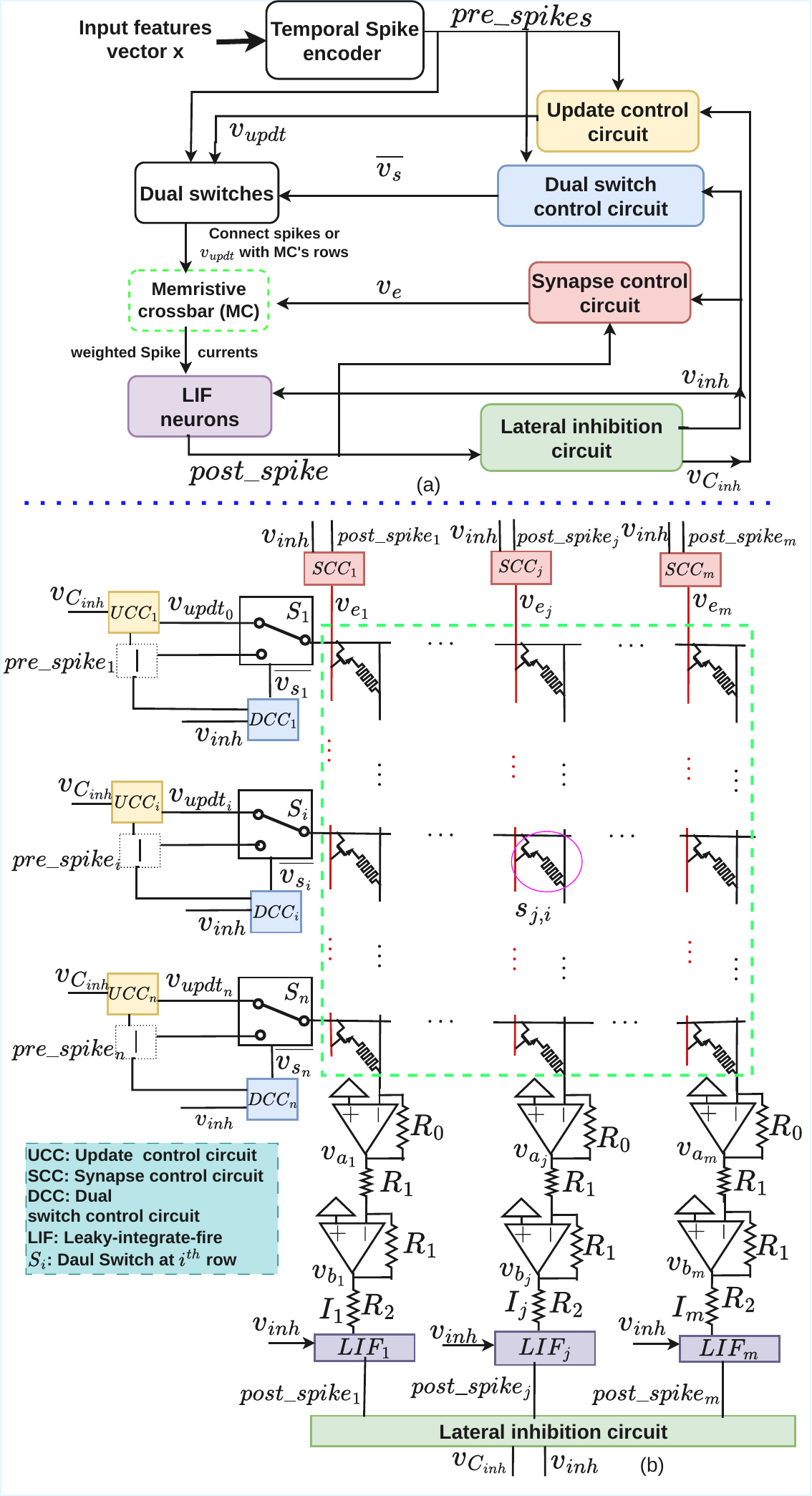}
    \caption{(a) Overview of the memristive SNN architecture. The arrows indicate the information flow. (b) Architectural layout of the memristive SNN. Switch $S_{i}$ enables $pre\_spike_{i}$ on the $i^{th}$ row. The $pre\_spike_{i}$ reaches post-neuron $LIF_{j}$ via synapse $s_{j,i}$ to contribute to integration and fire activities.}
    \label{fig:overview_architecture}
\end{figure}

Fig.~\ref{fig:overview_architecture}(a) illustrates the overview of the proposed memristive SNN architecture. The main components include a temporal spike encoder, dual switches, a memristive crossbar (MC), Leaky Integrate-and-Fire (LIF) neurons, and various proposed control modules such as the lateral inhibition circuit (LIC), synapse control circuit (SCC), dual switch control circuit (DCC), and update control circuit (UCC). The arrows indicate the flow of signals among these components.
The working of memristive SNNs is as follows:
\begin{itemize}
    \item The input feature vector $\mathbf{x}$ is pre-processed and then converted into $pre\_spikes$ using a temporal spike encoder.

\item These spikes are sent to the dual switches, dual switch control circuits, and update control circuits.

\item The dual switches transmit either $pre\_spikes$ or synapse update voltages ($v_{updt}$) to the MC, depending on the control signal $\mathbf{\overline{v_{s}}}$.

\item Within the MC, two primary operations take place: the production of weighted spike currents and the modification of synaptic weights (conductance of memristors).

\item The LIF neurons receive these weighted spike currents to generate a $post\_spike$. 

\item This $post\_spike$ is then sent to both the LIC and SCC modules.

\item In response to the $post\_spike$, the LIC generates signals $v_{inh}$ and $v_{C_{inh}}$.

\item The UCC utilizes $v_{C_{inh}}$ to produce $v_{updt}$. 

\item The $v_{inh}$ signal is forwarded to the LIF neurons, SCC, and DCC to perform lateral inhibition and establish a refractory period, generating $v_{e}$, and $\overline{v_{s}}$, respectively. The $v_{e}$ and $\overline{v_{s}}$ signals regulate the synapses and dual switches, respectively.
\end{itemize}

The layout of the memristive SNN architecture is shown in Fig.~\ref{fig:overview_architecture}(b). For a dataset with $n$ features and $m$ classes, it requires a 1T1M memristor crossbar of size $n \times m$, as indicated by the dashed green box in Fig.~\ref{fig:overview_architecture}(b). This setup includes $n$ pre-spikes, one for each feature, and $m$ post-neurons correspond to $m$ classes. Each $i^{th}$ row in the crossbar is associated with an Update Control Circuit ($UCC_{i}$), a Dual Switch Control Circuit ($DCC_{i}$), and a dual switch $S_{i}$. The $j^{th}$ column has a Synapse Control Circuit ($SCC_{j}$) at the top and two operational amplifiers (Opamps) and a Leaky Integrate-and-Fire neuron ($LIF_{j}$) at the bottom. In the $i^{\text{th}}$ column, $R_0$ serves as the feedback resistor of the first operational amplifier, producing the output voltage $v_{a_j}$. This voltage is applied to the second operational amplifier through the resistor $R_1$, generating the output voltage $v_{b_j}$. The second operational amplifier employs a feedback resistor equal to $R_1$. The voltage $v_{b_j}$ is then applied to $LIF_j$ through $R_2$, resulting in the input current $I_j$ for the neuron. Two cascaded inverting operational amplifiers are used to achieve an overall non-inverting (positive) gain. The 1T1M synapse employs the memristor model described in Section~\ref{memristor}. The boundary resistances of the memristor are denoted by $R_{on}$ and $R_{off}$ (see~\autoref{dynamic}). All LIF neurons are connected to a lateral inhibition circuit.

With this overview and architectural layout, we describe each component and its functionalities, illustrating spice simulations of a memristive SNN with 12 pre-spikes and three post-neurons.

\subsection{Temporal Spike Encoder}
\label{temp}
The inputs to the memristive SNN are spikes. Therefore, each element $x_{i}$ of vector $\mathbf{x}$ is encoded into spikes and spread into a time sequence length $T$. The temporal spike encoder is a functional block that converts $x_{i}$ into $pre\_spike_{i}$. In this work, the temporal spike encoding scheme is used where $pre\_spike_{i}$ corresponding to feature $x_{i}$ arrives at post neurons at the moment ($T$-$x_{i}$)~\cite{sboev2021mathematics,sboev2020solving}.

For simulation, the pre-spikes are represented with triangular pulses with $1\mu Sec$ rise and fall times. The height of pre-spikes is $1.1~V$ within the memristors' thresholds. The simulated pre-spikes are presented in Figs.~\ref{fig:post_spikes}(a). The $pre\_spike_{i}$ is given to dual switch $S_{i}$ to pass it to the $i^{th}$ row of MC.

\subsection{Dual Switches}
\label{dual}
As shown in Fig.~\ref{fig:overview_architecture}(b), the $i^{th}$ row of MC is associated with a dual switch $S_{i}$. The $S_{i}$ either allows the update voltage $v_{{updt}_{i}}$ from $UCC_{i}$ or the $pre\_spike_{i}$ to the $i^{th}$ row, depending on the switch control signal $\overline{v_{{s}_{i}}}$ coming from $DCC_{i}$. These voltage-controlled dual switches may be designed in the Spice simulator~\cite{li2022nca}.

\subsection{Memristive Crossbar}
The memristive crossbar employed here is of the 1T1M type as shown in Fig.~\ref{fig:overview_architecture}(b), green dashed box. There are $n$ rows and $m$ columns corresponding to the number of pre\_spikes and output neurons, respectively. Each artificial synapse $s_{j, i}$ connecting the $i^{th}$ row and $j^{th}$ column, as shown in the pink circle in Fig.~\ref{fig:overview_architecture}(b), consists of one transistor and one memristor. The synapses of $j^{th}$ column are made active or inactive by synapse control voltage $v_{e{_j}}$ from $SCC_{j}$. The MC represents the synaptic weight matrix where conductance $G_{j, i}$ of memristor $M_{j, i}$ representing weight $w_{j, i}$ of synapse $s_{j, i}$ between $pre\_spike_{i}$ and post-neuron $LIF_{j}$ is a matrix element. On arrival of $pre\_spike_{i}$ at $s_{j, i}$, it produces a weighted spike current proportional to $G_{j, i}$. The $j^{th}$ column with the conjunction of two operational amplifiers outputs current $I_{j}$ proportional to the sum of spike currents of synapses $s_{j, i}$, where $i=1~to~n$. The current $I_{j}$ is passed to $LIF_{j}$ to fire the $post\_spike_{j}$.

\subsection{Leaky-Integrate-Fire Neuron}
\begin{figure}[h]
    \centering
    \includegraphics[width=0.8\linewidth]{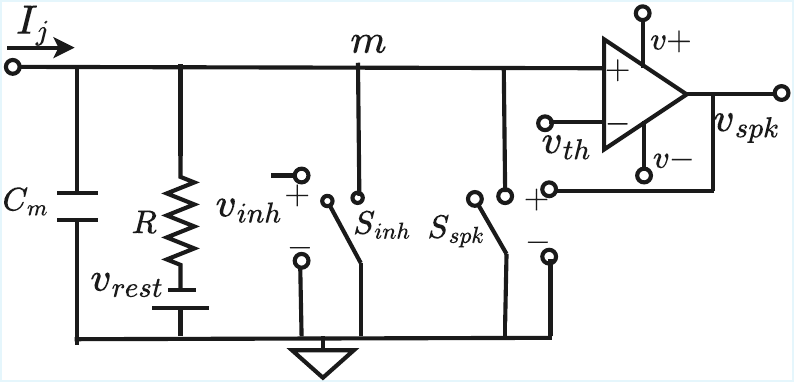}
    \caption{The Leaky-Integrate-Fire (LIF) schematic. The $I_{j}$ is the weighted spike current for $LIF_{j}$ from the $j^{th}$ column of MC. The circuit parameters: $C_{m}=5\mu F$, $R=50k\Omega$, $v_{rest}=0V$, the threshold of switches $S_{inh}$ and $S_{spk}$ are $1.5V$ and $0.45V$ respectively, $v_{th}=1mV$, $v+$ and $v-$ are $1V$ and $0V$ respectively.}
    \label{fig:lif}
\end{figure}

Fig.~\ref{fig:lif} shows the proposed LIF neuron schematic. It consists of membrane capacitor $C_{m}$, leaking resistor $R$, resting potential $v_{rest}$, and two voltage-controlled switches named $S_{inh}$ and $S_{spk}$ controlled by $v_{inh}$ and $v_{spk}$ respectively. The weighted spike current $I_{j}$ coming from column $j$ of MC is integrated across $C_{m}$ in the $LIF_{j}$ circuit. The $v_{C_{m}}$ increases gradually because of this integration. The Opamp compares threshold potential $v_{th}$ and membrane potential $v_{C_{m}}$ across capacitor $C_{m}$. When $v_{C_{m}}<v_{th}$ then voltage $v_{spk}$ is $v-$ but when $v_{C_{m}}$ starts exceeding $v_{th}$, the $v_{spk}$ suddenly starts increasing from $v-$ to $v+$. As soon as $v_{spk}$ crosses the threshold of switch $S_{spk}$, which is less than $v+$, then $S_{spk}$ becomes closed, resulting in the sudden discharging of $C_{m}$, and $v_{spk}$ again comes back to $v-$. The shape of $v_{spk}$, generated because of the sudden increase and decrease of $v_{spk}$, is termed as $post\_spike_{j}$.

A resistor \( R \) connected to the resting potential \( v_{rest} \) provides a path for charge leakage. The inhibitory switch $S_{inh}$ controlled by inhibitor pulse $v_{inh}$ is employed to implement lateral inhibition and the refractory period. If $v_{inh}$ is at peak voltage, then it closes $S_{inh}$, resulting in sudden discharge of $C_{m}$ as well as no charge integration across $C_{m}$, and no spike till the width of $v_{inh}$. So here the width of $v_{inh}$ is called the refractory period.

\begin{figure}[h]
    \centering
    \includegraphics[width=0.8\linewidth]{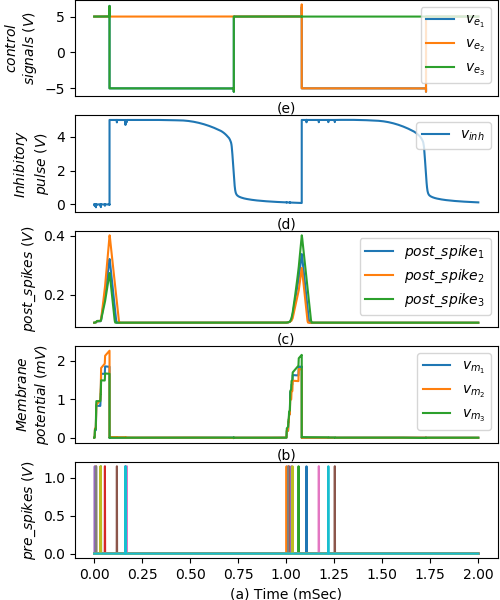}
    \caption{(a) Encoded pre-spikes. (b) The membrane potential dynamics of three LIF neurons in response to pre-spikes' stimulus. (c) The $post\_spike_{2}$ and $post\_spike_{3}$ are the post-spikes of winner neurons $LIF_{2}$ and $LIF_{3}$. (d) The inhibition pulses $v_{inh}$. (e) The behavior of synapse control signal $v_e{_{j}}$ for $LIF_{j}$, where j=1 to 3.}
    \label{fig:post_spikes}
\end{figure}

We simulated the LIF circuit, Fig.~\ref{fig:lif}, on the LTspice simulator, and the results are shown in Figs.~\ref{fig:post_spikes}(a) to (c). On arrival of pre-spikes, Fig.~\ref{fig:post_spikes}(a), the membrane potentials of three post-neurons are increasing gradually as shown in Fig.~\ref{fig:post_spikes}(b). In Fig.~\ref{fig:post_spikes}(c), $post\_spike_{2}$ and $post\_spike_{3}$ are fired by $LIF_{2}$ and $LIF_{3}$ when $v_{m_{2}}$ and $v_{m_{3}}$ cross $v_{th}$ respectively.

\subsection{Lateral Inhibition Circuit}
\label{LIC}
\begin{figure}[h]
    \centering
    \includegraphics[width=0.8\linewidth]{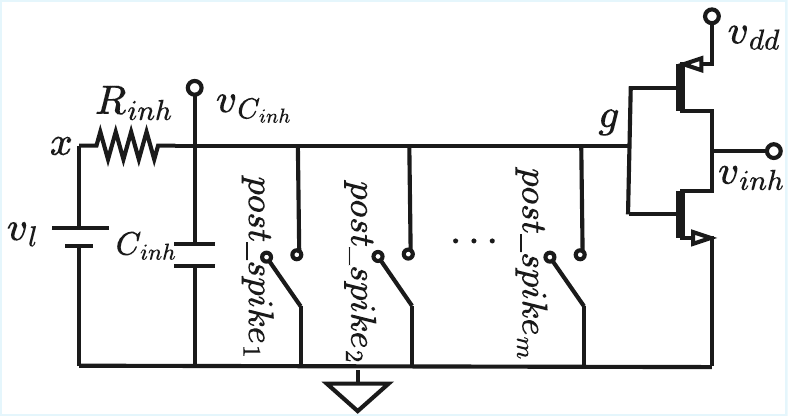}
    \caption{Lateral inhibition circuit (LIC) -- the $m$ parallel connected switches are controlled by respective post-spikes from $m$ LIFs. The circuit parameters are $v_{l}=5V$, $R_{inh}=1k\Omega$, $C_{inh}=1.5\mu F$, $v_{dd}=5~V$ and the threshold of switches are $0.4V$.}
    \label{fig:wta}
\end{figure}

The lateral inhibition circuit (LIC) is shown in Fig.~\ref{fig:wta}. This circuit consists of $m$ voltage-controlled switches controlled by post-spikes coming from $m$ LIFs, respectively. Additionally, there are voltage source $v_{l}$, $R_{inh}$, a capacitor $C_{inh}$, and a CMOS. The $C_{inh}$ is charged by $v_{l}$ through $R_{inh}$. Initially, $C_{inh}$ is fully charged and the amplitude of $v_{inh}$ is low, i.e., at base voltage. The occurrence of the $post\_spike_j$ closes the corresponding switch that suddenly discharges $C_{inh}$, making $v_{inh}$ high, i.e., at peak voltage until $C_{m}$ is recharged again by $v_{l}$. In this way, the voltage pulse $v_{inh}$ is generated.

The $v_{inh}$ is input to all LIFs. Whenever any LIF fires a spike, suddenly $v_{inh}$ from LIC closes $S_{inh}$ switches of all LIFs that make discharging of their membrane capacitors. It ensures that no more spikes from any of the LIFs. Therefore, the one who spikes first is the winner here and thus performs the WTA operation. There is no integration across $C_{m}$ of any LIFs till the width of $v_{inh}$. The width of $v_{inh}$ decides the refractory period.

In the LIC schematic of Fig.~\ref{fig:wta}, the voltage across $C_{inh}$, denoted as $v_{C_{inh}}$, follows RC dynamics governed by KCL. During charging, $C_{inh}\frac{dv_{C_{inh}}}{dt}=\frac{v_l-v_{C_{inh}}}{R_{inh}}$. When any post-synaptic neuron fires, the corresponding switch turns ON, introducing a discharge path modeled by a switching function $S(t)\in\{0,1\}$, where $S(t)=1$ during a post-spike. The overall dynamics are expressed as
\begin{equation}
C_{inh}\frac{dv_{C_{inh}}}{dt}=\frac{v_l-v_{C_{inh}}}{R_{inh}}-S(t)\frac{v_{C_{inh}}}{R_{switch\_on}},
\end{equation}
where $R_{switch\_on}$ is the ON resistance of the activated switch. The inhibition voltage generated through a CMOS inverter  driven by $v_{C_{inh}}$ is
\begin{equation}
v_{inh}=
\begin{cases}
v_{DD}, & v_{C_{inh}}<v^{inverter}_{th}\\
0, & \text{otherwise}.
\end{cases}
\end{equation}
Thus, a post-spike discharges $C_{inh}$, reducing $v_{C_{inh}}$ and generating a high $v_{inh}$ that inhibits all neurons, thereby enforcing WTA operation. The inhibition pulse width is determined by the RC time constant $\tau=R_{inh}C_{inh}$, corresponding to the refractory period.

The simulation results in Figs.~\ref{fig:post_spikes}(a) to (d) of LIC and LIF explain the steps of $v_{inh}$ generation, refractory period, and WTA process. The pre-spikes, as shown in Fig.~\ref{fig:post_spikes}(a), increase the membrane potentials of LIFs shown in Fig.~\ref{fig:post_spikes}(b). The $v_{m_{2}}$ and $v_{m_{3}}$,  of $LIF_{2}$ and $LIF_{3}$ crosses $v_{th}$, in Fig.~\ref{fig:post_spikes}(b), and they fire spikes as shown in Fig.~\ref{fig:post_spikes}(c) that causes to generate $v_{inh}$ pulses shown in Fig.~\ref{fig:post_spikes}(d). It may be noticed that there is no further rise in any membrane potential despite the arrival of pre-spikes till the refractory period or the width of pulse $v_{inh}$.

\subsection{Synapse Control Circuit}
\begin{figure}[h]
    \centering
    \includegraphics[width=0.9\linewidth]{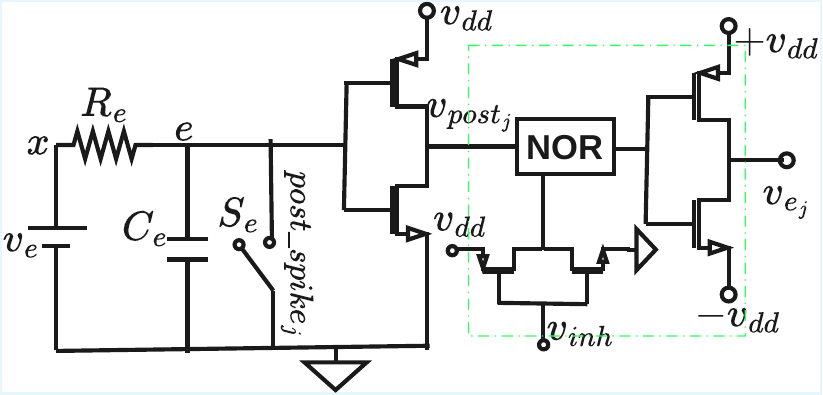}
    \caption{The schematic of the synapse control circuit. The circuit parameters: $v_{e}=5V$, $R_{e}=1k\Omega$, $C_{e}=1.7\mu F$, $v_{dd}=5V$, $+v_{dd}=5V$, $-v_{dd}=-5V$, and threshold of switch $S_{e}$ is $0.4V$.}
    \label{fig:synapse}
\end{figure}

The employed 1T1M synapse becomes active when transistor T is ON and inactive when T is OFF. There are two scenarios that happened in MC; before the post-spike from the winner LIF, all synapses generate a weighted spike current on the arrival of pre-spikes, but after the firing of the post-spike, the synapses of only the winner neuron are updated. In the first case, all synapses are active, whereas in the second case, only the synapses of the winner neuron are active and the rest are inactive. Supporting the above logic, the synapse control circuit (SCC) of $LIF_{j}$ in Fig.~\ref{fig:synapse} generates a voltage signal $v_{e_{j}}$. The $v_{e_{j}}$ turns T ON or OFF all synapses of $LIF_{j}$ if it is at peak or base voltage levels, respectively. In other words, the amplitude of $v_{e_{k}}$, for all $k=1~to~m$, is high before the winner's post-spike and thereafter only $v_{e_{j}}$ is at peak if $LIF_{j}$ is the winner.

Initially, the capacitor \( C_e \) is fully charged, and the voltage \( v_{post_j} \) is low. When \( post\_spike_j \) occurs, switch \( S_e \) closes, \( C_e \) discharges, and the amplitude of \( v_{post_j} \) becomes high. The signal \( v_{e_j} \) remains low only if \( LIF_j \) has not spiked while any other LIF neuron has. This behavior of $v_{e_{j}}$ corresponds to the logical operation \( \overline{v_{e_j}} = \overline{v_{post_j}} \wedge v_{inh} \), where \( v_{inh} \) is the inhibition signal coming from LIC module. The green dotted section of the schematic in Fig.~\ref{fig:synapse} implements this logic. For all non-winner neurons $LIF_{k}$, the amplitude of \( v_{e_k} \) is at a low level, turning the synapses connected to \( LIF_k \) OFF, thereby preventing unwanted weight modification. The NOR block in this work is implemented with MOSFET transistors. 

To mathematically describe the SCC in Fig.~\ref{fig:synapse}, the voltage across $C_e$, denoted as $v_{C_e}$, follows RC dynamics. During normal operation, $C_e\frac{dv_{C_e}}{dt}=\frac{v_e-v_{C_e}}{R_e}$. When the post-spike signal $post\_spike_j$ occurs, the switch $S_e$ turns ON, introducing a discharge path modeled by a switching function $S_j(t)\in\{0,1\}$, where $S_j(t)=1$ during the spike. The overall dynamics are given by
\begin{equation}
C_e \frac{dv_{C_e}}{dt} = \frac{v_e - v_{C_e}}{R_e} - S_j(t)\frac{v_{C_e}}{R_{switch\_on}},
\end{equation}
where $R_{switch\_on}$ is the ON resistance of the switch. The resulting voltage is converted into a digital-like signal $v_{post_j}$ through a CMOS inverter:
\begin{equation}
v_{post_j} =
\begin{cases}
v_{dd}, & v_{C_e}<V^{inverter}_{th}\\
0, & \text{otherwise}.
\end{cases}
\end{equation}
The control signal $v_{e_j}$ follows $\overline{v_{e_j}}=\overline{v_{post_j}}\wedge v_{inh}$, equivalently $v_{e_j}=v_{post_j}\lor\overline{v_{inh}}$, and is expressed as
\begin{equation}
v_{e_j} =
\begin{cases}
v_{dd}, & v_{post_j}=v_{dd}\ \text{or}\ v_{inh}=0\\
-\,v_{dd}, & \text{otherwise}.
\end{cases}
\end{equation}
Thus, before inhibition ($v_{inh}=0$), all synapses remain active, whereas after inhibition, only the neuron generating the post-spike maintains $v_{e_j}$ at a high level, enabling selective synaptic update for the winner neuron.

In Fig.~\ref{fig:post_spikes}(e), all $v_{e_{1}}$, $v_{e_{2}}$, and $v_{e_{3}}$ are high before post spikes in Fig.~\ref{fig:post_spikes}(c) but after $post\_spike_{2}$ and $post\_spike_{3}$ only $v_{e_{2}}$ and $v_{e_{3}}$ are continued to be high till the width of $v_{inh}$.

\subsection{Dual Switch Control Circuit}
\begin{figure}[h]
    \centering
    \includegraphics[width=0.8\linewidth]{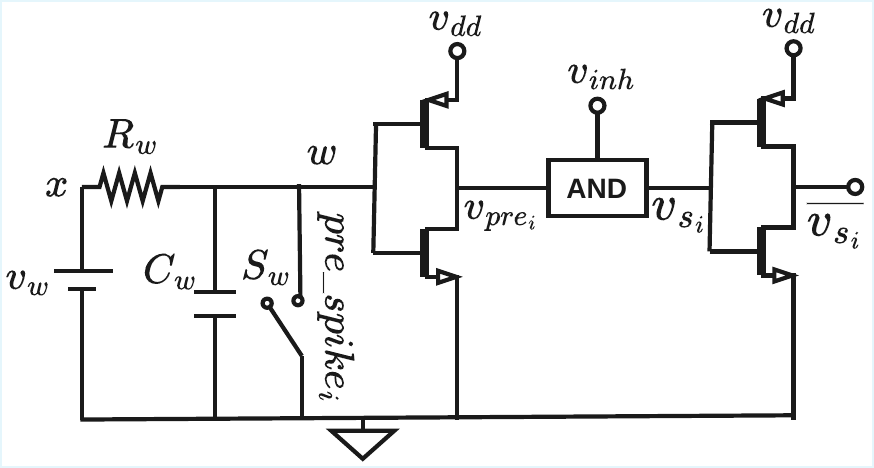}
    \caption{The dual switch control circuit. The circuit parameters: $v_{w}=5V$, $C_{w}=1.5\mu F$, $R_{w}=1k\Omega$, $v_{dd}=5V$, and threshold of switch $S_{w}$ is $1.1V$.}
    \label{fig:switch}
\end{figure}

Either $pre\_spike_{i}$ or voltage $v_{updt_{i}}$ is allowed to $i^{th}$ row of MC at a time by dual switch $S_{i}$ (refer Fig.~\ref{fig:overview_architecture}(b)). The $v_{updt_{i}}$ updates active synapse of row $i$. This $S_{i}$ is controlled by voltage $\overline{v_{s_{i}}}$ coming from the $DCC_i$, shown in Fig.~\ref{fig:switch}, present at $i^{th}$ row. When $\overline{v_{s_{i}}}$ is at peak then $S_{i}$ allows $pre\_spike_{i}$ else $v_{updt_{i}}$. When both $pre\_spike_{i}$ has occurred and the winner post-neuron $LIF_{j}$ has fired a $post\_spike_{j}$, not necessarily at the same time, then only the $S_{i}$ connects $v_{{updt}_{i}}$ to the $i^{th}$ row of the MC to update memristor $M_{j,i}$ otherwise it allows $pre\_spike_{i}$.

In Fig.~\ref{fig:switch} representing DCC at $i^{th}$ row, the switch \(S_w\) is controlled by \(pre\_spike_i\). Initially, the capacitor \( C_w \) is fully charged by $v_{w}$ through $R_{w}$, and the voltage \( v_{pre_i} \) is low. When \(pre\_spike_i\) occurs, it turns \(S_w\) ON, discharging capacitor \(C_w\) and raising \(v_{pre_i}\). When both \(v_{pre_i}\) and \(v_{inh}\) are at high levels, then this indicates the occurrence of \(pre\_spike_i\) and the post-spike of the winner neuron, respectively. Consequently, the circuit generates a low (base) voltage \(\overline{v_{s_i}}\), signaling the start of synaptic weight update. The time duration till \(\overline{v_{s_i}}\) is low, which denotes the update period of synapses $s_{j,i}$ of the winner neuron, $LIF_{j}$. The AND block is implemented with MOSFET transistors.

To mathematically describe the DCC in Fig.~\ref{fig:switch}, the voltage across $C_w$, denoted as $v_{C_w}$, follows RC dynamics. During normal operation, $C_w\frac{dv_{C_w}}{dt}=\frac{v_w-v_{C_w}}{R_w}$. When the pre-synaptic spike $pre\_spike_i$ occurs, the switch $S_w$ turns ON, introducing a discharge path modeled by a switching function $S_i(t)\in\{0,1\}$, where $S_i(t)=1$ during the spike. The overall dynamics are expressed as
\begin{equation}
C_w \frac{dv_{C_w}}{dt} = \frac{v_w - v_{C_w}}{R_w} - S_i(t)\frac{v_{C_w}}{R_{switch\_on}},
\end{equation}
where $R_{switch\_on}$ is the ON resistance of the switch. The resulting voltage is converted into a digital-like signal $v_{pre_i}$ through a CMOS inverter:
\begin{equation}
v_{pre_i} =
\begin{cases}
v_{dd}, & v_{C_w}<V^{inverter}_{th}\\
0, & \text{otherwise}.
\end{cases}
\end{equation}
The control signal $\overline{v_{s_i}}$ follows the logical relation $\overline{v_{s_i}}=v_{pre_i}\wedge v_{inh}$ and is expressed as
\begin{equation}
\overline{v_{s_i}} =
\begin{cases}
v_{dd}, & v_{pre_i}=v_{dd}\ \text{and}\ v_{inh}=v_{dd}\\
0, & \text{otherwise}.
\end{cases}
\end{equation}
Thus, the dual switch $S_i$ connects $v_{updt_i}$ to the $i^{th}$ row only when both the pre-synaptic spike and inhibition signal are present, indicating that a winner neuron has fired; otherwise, $pre\_spike_i$ is propagated, ensuring selective synaptic update during learning while preserving normal spike transmission during inference.

\begin{figure}[h]
    \centering
    \includegraphics[width=0.9\linewidth]{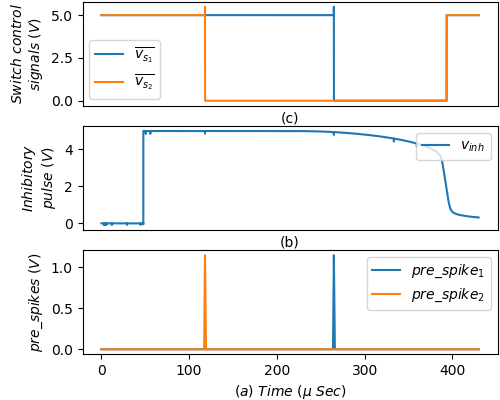}
    \caption{(a) The features encoded into $pre\_spikes$. (b) The start of the inhibition pulse $v_{inh}$ from LIC in Fig.~\ref{fig:wta} indicates the occurrence of the winner $post\_spike_{j}$. (c) The $\overline{v_{s_{i}}}$ allows the $v_{updt_{i}}$ to modify synapse $s_{j,i}$ when it is low. The amplitude of $\overline{v_{s_{i}}}$ is low i.e., the $s_{j,i}$ indicating both $pre\_spike_{i}$ and $post\_spike_{j}$ have occurred.}
    \label{fig:prespikes}
\end{figure}

The spice simulation results of DCCs of $row_{1}$ and $row_{2}$ are illustrated in Figs.~\ref{fig:prespikes}(a) to (c). In Fig.~\ref{fig:prespikes}(a) two pre-spikes at $row_{1}$ and $row_{2}$ are occurred. When the $v_{inh}$ becomes high in Fig.~\ref{fig:prespikes}(b), it indicates the timing of the occurrence of the post-spike. In Fig.~\ref{fig:prespikes}(c), the $\overline{v_{s_{1}}}$ and $\overline{v_{s_{2}}}$ are low when the winner neuron fired the post-spike and two pre-spikes ($pre\_spike_{1}$ and $pre\_spike_{2}$) have occurred. Therefore, the $v_{updt_{1}}$ and $v_{updt_{2}}$ are allowed by $S_{1}$ and $S_{2}$ to $row_{1}$ and $row_{2}$ till the $\overline{v_{s_{1}}}$ and $\overline{v_{s_{2}}}$ are low respectively.

\subsection{Update Control Circuit}
\label{ucc}
\begin{figure}[h]
    \centering
    \includegraphics[width=0.8\linewidth]{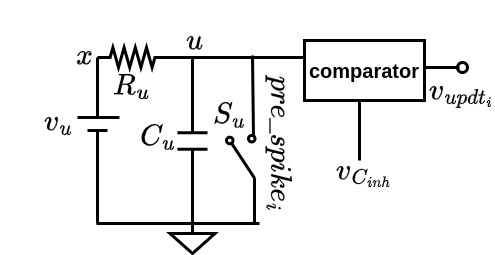}
    \caption{The schematic of the update control circuit. The circuit parameters: $v_{u}=5V$, $C_{u}=1.5\mu F$, $R_{u}=1k\Omega$, threshold of $S_{u}$ is $1.1V$. if \(v_{C_u} > v_{C_{inh}}\) then $v_{updt_{i}}=1.4V$ else $v_{updt_{i}}=-2.6V$.}
    \label{fig:update}
\end{figure}

The conductance of the memristor in an active synapse is increased or decreased by applying positive or negative voltages across it, respectively. For a synapse $s_{j,i}$, the synaptic weight i.e., conductance $G_{j,i}$ is increased if $pre\_spike_{i}$ occurs before $post\_spike_{j}$ and vice-versa. The UCC in Fig.~\ref{fig:update} produces $v_{updt_{i}}$ to modify the $G_{j, i}$ depending on the temporal order of occurrence of these spikes. In order to increase \(G_{j, i}\) of memristor \(M_{j, i}\) in synapse \(s_{j, i}\), the voltage \(v_{updt_i}\) must be positive and greater than the memristor's threshold voltage. Conversely, to decrease \(G_{j, i}\), the opposite conditions must hold. In the LIC schematic in Fig.~\ref{fig:wta}, the $C_{inh}$ is discharged by the winner's post-spike, and in Fig.~\ref{fig:update}, $C_{u}$ is discharged by $pre\_spike_{i}$. These capacitors are recharged by $v_l$ and $v_{u}$ through $R_{inh}$ and $R_{u}$ respectively where $v_l=v_{u}$ and $C_{inh}=C_{u}$. By comparing the voltages $v_{C_{inh}}$ and $v_{C_{u}}$ across $C_{inh}$ and $C_{u}$ respectively, the order of occurrence of $pre\_spike_{i}$ and $post\_spike_{j}$ is predicted. The comparator compares \(v_{C_{inh}}\) with \(v_{C_u}\) to produce \(v_{updt_i}\). If \(pre\_spike_i\) occurs before \(post\_spike_j\), then \(v_{C_u} > v_{C_{inh}}\), resulting in a positive \(v_{updt_i}\). Conversely, if \(pre\_spike_i\) occurs after \(post\_spike_j\), then \(v_{C_u} < v_{C_{inh}}\), and \(v_{updt_i}\) is negative.

\begin{figure}[h]
    \centering
    \includegraphics[width=0.8\linewidth]{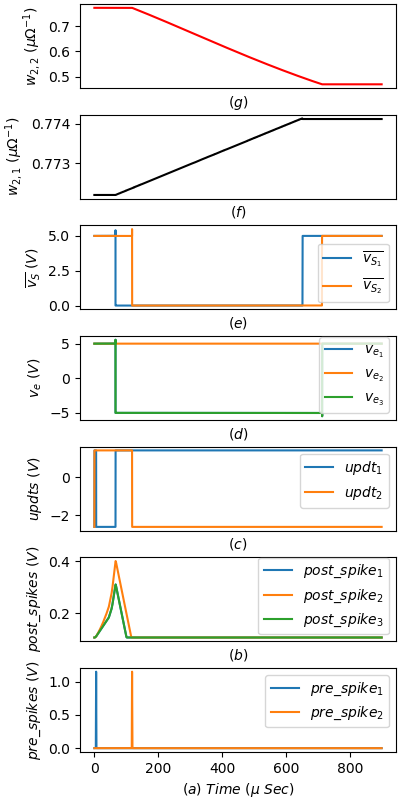}
    \caption{The plots (a) to (g) are to explain training steps. (a) pre-spikes. (b) $LIF_{2}$ is winner firing $post\_spike_{2}$. (c) Update voltages in rows 1 and 2. (d) Only $v_{e_{2}}$ is high during the update, making synapses of $LIF_{2}$ active. (e) The update windows for synapses $s_{2,1}$ and $s_{2,2}$ are decided by $\overline{v_{s_{1}}}$ and $\overline{v_{s_{2}}}$ being low. The weights $w_{2,1}$ and $w_{2,2}$ in plots (f) and (g) increased and decreased following the orders of pre- and post-spikes.}
    \label{fig:updt}
\end{figure}

The simulated outputs of UCC are shown in Figs.~\ref{fig:updt}(a) to (c). In Fig.~\ref{fig:updt}(a), $pre\_spike_{1}$ and $pre\_spike_{2}$ occurred before and after $post\_spike_{2}$ in Fig.~\ref{fig:updt}(b). The $v_{updt_{1}}$ and $v_{updt_{2}}$ are produced by $UCC_{1}$ and $UCC_{2}$ respectively, in Fig.~\ref{fig:updt}(c). In the proposed model, a synapse starts updating its weight only when both pre- and post-spikes have occurred. Therefore, for the synapse $s_{j,i}$ the value of $v_{updt_{i}}$ is considered effective only when $pre\_spike_{i}$ and $post\_spike_{j}$ have occurred.

\section{Supervised In-situ STDP Training}
\label{train}
The memristive SNN is trained using the proposed hardware-friendly supervised in-situ STDP algorithm. The in-situ training is capable of mitigating any hardware imperfections present in  devices~\cite{li_2018_nature_communication}, and it is more energy-efficient and faster due to reduced communication with external hardware~\cite{alibart_2013_nature_communication}. The dataset is divided into training and test sets. Training is performed with the training set, while testing is done over the test set. The input features are converted into spikes as explained in Sections \ref{temp}. The training involves iteratively presenting input spikes from the training set to the memristive SNN using the procedure outlined in~\autoref{algo:weightUpdate}. Here, the data sample's label supervises the training procedure.

According to STDP rules for synapse \(s_{j,i}\), if the \(pre\_spike_{i}\) occurs before the \(post\_spike_{j}\), the conductance \(G_{j,i}\) ($\equiv w_{j,i}$) of the memristor $M_{j,i}$ increases and vice-versa. The magnitude of the conductance change decreases with the increase in time difference between these spikes.

Initially, all memristors in the MC are set to a high conductance state \(G_{max}\). For an input with label $j$, only $LIF_{j}$ receives an extra bias current $I_{b_{j}}$ to achieve supervised training. The $I_{b_{j}}$ should be large enough to make $LIF_{j}$ the winner. The synaptic weight update begins when \(post\_spike_{j}\) occurs. If synapse \(s_{j,i}\) is ON (i.e., the amplitude of \(v_{e_{j}}\) is high) and the update voltage \(v_{updt_{i}}\) is applied to row \(i\) by making \(\overline{v_{{s}_{i}}}\) low, then the memristor \(M_{j,i}\) receives the \(v_{updt_{i}}\) across it that results in modification in \(G_{j,i}\) depending on the sign and magnitude of $v_{updt_{i}}$. All the synaptic weights of the winner neurons are updated in parallel. The update rules are detailed in Table~\ref{tab:update_rules}. The same \autoref{algo:weightUpdate} is used for both binary pattern recognition and input vector classification problems.
\begin{table}[h]
    \centering
    \caption{Update rules for synapse \(s_{j,i}\). *: don't care}
    \resizebox{\linewidth}{!}{%
    \begin{tabular}{|c|c|c|c|c|}
    \hline
        \(\overline{v_{{s}_{i}}}\) & \(v_{{e}_{j}}\) & \(pre\_spike_{i}\) & \(post\_spike_{j}\) & Weight \(G_{j,i}\) \\ \hline
        High & * & * & * & No change \\ \hline 
        * & Low & * & * & No change \\ \hline
        Low & High & Before & After & \(\uparrow\) \\ \hline
        Low & High & After & Before & \(\downarrow\) \\ \hline
    \end{tabular}}
    \label{tab:update_rules}
\end{table}

\begin{algorithm}[]
    \caption{Supervised In-situ STDP Training.
    }
    \label{algo:weightUpdate}
    \KwData{Temporal encoded spikes corresponding to input vector $\mathbf{x}$ with label \(j\);}
    \KwResult{The proper weight adjustment of all synapses of winner \(LIF_{j}\);}
     Make all synapses in the MC ON\;

    Allow \(pre\_spike_{i}\) corresponding to $x_{i}$, the $i^{th}$ feature of input $\mathbf{x}$, to $UCC_{i}$, $DCC_{i}$, and dual switch $S_{i}$ where \(i=1\) to \(n\)\;
     
    Apply an extra bias current \(I_{b_j}\) only to \(LIF_{j}\), ensuring supervised training\;
    
    \(LIF_{j}\) fires \(post\_spike_{j}\) first due to the sufficient bias current \(I_{b_j}\)\;

    The \(post\_spike_{j}\) is input to lateral inhibition circuit and all SCCs\;

    The \(post\_spike_{j}\) causes lateral inhibition circuit to output inhibitory pulse \(v_{inh}\) and voltage $v_{C_{inh}}$ across capacitor $C_{inh}$ in this circuit\;
    
    The \(v_{inh}\) is input to all LIFs, DCCs, and SCCs \;

    The \(v_{C_{inh}}\) is input to all UCCs\;
    
    The \(v_{inh}\) inhibits all \(LIF\) neurons from firing except $LIF_{j}$, making \(LIF_{j}\) the winner\;
    
    $SCC_{j}$ makes all the synapses ON connected to \(LIF_{j}\) while those connected to other neurons OFF\;
    
    Switch $S_{i}$ connects row \(i\) of MC to \(v_{updt_{i}}\) from $UCC_{i}$ during update time window\;
    
    Update \(G_{j,i}\) of memristor \(M_{j,i}\) in synapse \(s_{j,i}\) based on whether \(v_{updt_{i}}\) is positive (increase) or negative (decrease).
\end{algorithm}

\subsubsection{Pattern Recognition}
The labeled binary patterns of size $r \times c$, where r and c are the numbers of rows and columns, respectively, are encoded into spikes using temporal encoding. The bits of binary patterns are encoded row-wise into spikes and given to the memristive SNN for training using \autoref{algo:weightUpdate}. Here, no pre-processing is required. The examples of $5 \times 3$ binary patterns are shown in Fig.~\ref{fig:input_output_patterns}(a).

\subsubsection{Classification Problem}
In classification problems, the pre-processing of inputs with the Gaussian receptive field method is done to smooth the population of spikes as explained in Section~\ref{preprocessing}. After pre-processing of input vectors, they are encoded into spikes and given to \autoref{algo:weightUpdate}.

\subsubsection{ Testing}
During the testing phase, all synapses are ON, and pre-synaptic spikes corresponding to the input are applied to the rows of the MC. In this phase, synaptic weights are not updated. The recognition or classification of input $\mathbf{x}$ with label $j$ is said to be correct if and only if \(LIF_{j}\) fires a spike.

\section{Simulation and results}
\label{simulation}

The functioning of the memristive SNN and in-situ training Algorithm~\ref{algo:weightUpdate} is verified by simulating them using the LTspice simulator running on an Ubuntu 20.04 LTS environment with an 8-core 1.6GHz Intel Core i5 processor and 8GB RAM. Its functionality is verified for pattern recognition and vector classification tasks. The simulations are performed to recognize binary images in Fig.~\ref{fig:input_output_patterns}(a) and to classify labeled datasets like IRIS~\cite{bcwiris:2019} and Breast Cancer Wisconsin (BCW)~\cite{bcwiris:2019}. For different data sets, the network structure and parameters are shown in \autoref{tab:network_parameters}. The simulation employs the memristor model described by \cite{Zhang_tcas_2017} with positive and negative thresholds as $1.2V$ and $-2.4V$, respectively. The positive and negative values of update voltage $v_{updt}$ are $1.4V$ and $-2.6V$, respectively. The pre-spikes are triangular pulses with \(1~\mu\text{s}\) rise and fall times, and amplitude of \(1.1\text{V}\). The memristor’s minimum conductance \(G_{min}=0.0167~\mu \Omega^{-1}\) and maximum conductance \(G_{max}=1~\mu \Omega^{-1}\).
\begin{table*}[]
\caption{Network parameters used in LTspice simulation of memristive SNNs}
\label{tab:network_parameters}
\resizebox{\linewidth}{!}{%
\begin{tabular}{|c|c|c|c|c|c|c|c|c|}
\hline
Dataset &
  \begin{tabular}[c]{@{}l@{}}\# feature (n) or pre-spikes\\ after pre-processing\end{tabular} &
  \begin{tabular}[c]{@{}l@{}} \# classes (m) \\ or post-spikes \end{tabular}&
  \begin{tabular}[c]{@{}l@{}}MC size\\ \# synapses \end{tabular}&
  $R_{0}~(\Omega)$ &
  $R_{1}~(\Omega)$ &
  $R_{2}~(\Omega)$ &
  $C_{m}$ &
  $I_{b}$ \\ \hline
IRIS                                                               & 12                   & 3 & 36  & 95k & 1k & 500 & $5 \mu$F   & $35~\mu$A  \\ \hline
\begin{tabular}[c]{@{}l@{}}BCW\end{tabular} & 90                   & 2 & 180 & 40k & 1k & 500 & $7 \mu$F   & $100~\mu$A \\ \hline
$5 \times 3$                                                       & 15, no pre-processing & 4 &  60   & 50k & 1k & 1k  & $0.3 \mu$F & $500~\mu$A \\ \hline
\end{tabular}}
\end{table*}

The circuit parameters voltages, resistances, and capacitances, were selected through iterative circuit-level simulations and empirical optimization. In particular, the RC time constants and voltage levels were adjusted to maintain synchronisation among spike generation, inhibition, and synaptic update operations, thereby ensuring stable functionality and signal integrity.

The proposed analog memristive computing architecture eliminates DAC/ADC interfaces and frequent charging/discharging of digital nodes, thereby reducing power consumption, area, and latency. It also enables in-memory (in-situ) computation within the memristive crossbar, mitigating the von Neumann memory bottleneck while naturally supporting parallel processing. Compared to prior works~\cite{zhao2018neurocomputing, zhang2021ScienceBulletin, li2021ICIST, li2022nca, jiang2022memristor, Liu_2024_tvlsi, Verma2025tcas}, the proposed architecture offers (i) a modular and scalable structure, (ii) in-situ supervised learning capability absent in several existing analog/mixed-signal designs, and (iii) reduced hardware complexity suitable for energy-constrained applications. Furthermore, Table~\ref{tab:comparison} (column 8) qualitatively compares the proposed architecture with existing analog and mixed-signal designs in terms of hardware complexity, functionality, and performance.

Fig.~\ref{fig:updt} illustrates the intermediate simulation results of the training of the memristive SNN. Figs.~\ref{fig:updt}(a) to (g) brief the synaptic plasticity process during the training of the memristive SNN. In Fig.~\ref{fig:updt}(a), $pre\_spike_{1}$ and $pre\_spike_{2}$ occur before and after $post\_spike_{2}$ of winner $LIF_{2}$, in Fig.~~\ref{fig:updt}(b), respectively. In Fig.~\ref{fig:updt}(c), the update voltages $v_{updt_{1}}$ and $v_{updt_{2}}$ from $UCC_{1}$ and $UCC_{2}$ are shown. Accordingly, only $v_{e_{2}}$ are high making all synapses of $LIF_{2}$ active and rest of the synapses are inactive as $v_{e_{1}}$ and $v_{e_{3}}$ are low, refer Fig.~\ref{fig:updt}(d). When both pre and post-spikes have occurred then only $\overline{v_{s_{1}}}$ and $\overline{v_{s_{2}}}$ are low, in Fig.~\ref{fig:updt}(e), that means the update voltages $v_{updt_{1}}$ and $v_{updt_{2}}$ are fed to $row_{1}$ and $row_{2}$ to MC respectively. In Figs.~\ref{fig:updt}(f) and (g) following the update rules in \autoref{tab:update_rules}, the weights $w_{2,1}$ and $w_{2,2}$ are adjusted based on the temporal ordering of $pre\_spike_{1}-post\_spike_{2}$ and $pre\_spike_{2}-post\_spike_{2}$. This demonstrates the correctness of the proposed training method.

\subsection{Simulation for pattern recognition}
\begin{figure*}[h]
    \centering
    \includegraphics[width=0.8\linewidth]{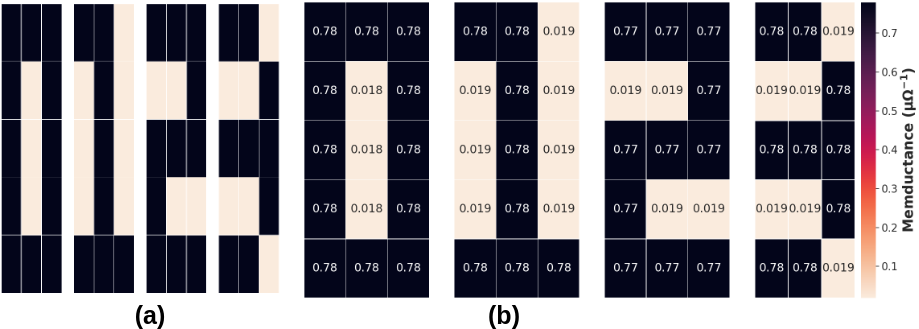}
    \caption{(a) Binary training patterns, and (b) heat map of final synaptic weights in $\mu \Omega^{-1}$, which are consistent after training with the patterns used.}
    \label{fig:input_output_patterns}
\end{figure*}

The four binary patterns of size \(5 \times 3\) labeled 0, 1, 2, and 3 are chosen to train the memristive SNN, as shown in Fig.~\ref{fig:input_output_patterns}(a). There are $15$ pre-neurons and four post-neurons.
Temporal spike encoding is used to map the pixels into pre-spikes, where a pre-spike corresponding to a black pixel occurs \(0.5~\text{ms}\) {\it before} than that of the white pixel.
The training set comprises \(100\) samples (patterns 0, 1, 2, and 3 with 25 each). After training, the memristive SNN recognizes all four patterns correctly in the testing phase. The final synaptic weights are visualized as a heat map in Fig.~\ref{fig:input_output_patterns}(b), demonstrating the consistency with the input patterns.

\subsection{Robustness Against Noisy Patterns}
For pattern recognition tasks, the robustness of our memristive SNN was evaluated by introducing noise in the patterns through pixel flips (black to white or white to black). Our approach demonstrated better noise resilience compared to previous literature \cite{li2021ICIST, li2022nca,zhou_iscas_2022}, as shown in Table~\ref{tab:noise}.
\begin{table}[h]
    \centering
    \caption{Average recognition accuracy under different noise levels.}
    \resizebox{\linewidth}{!}{%
    \begin{tabular}{|c| c| c| c| c|} \hline
        Noise level  & In~\cite{li2021ICIST} & In~\cite{li2022nca} & In~\cite{zhou_iscas_2022}  & In proposed work \\ \hline
        noiseless & 100\% & 100\% & 97.5\%  & 100\%\\
        6.67\%  &96.67\% & 93.33\% & -------& 100\%\\
        13.33\%  & 87.46\% & 76.9\% & -------& 95.47\%\\
        20\% & --------& ------ & -------& 93.4\% \\
        \hline
    \end{tabular}}
    \label{tab:noise}
\end{table}

\subsection{Scalability of the Memristive SNN architecture}
\begin{figure}[h]
    \centering
    \includegraphics[width=\linewidth]{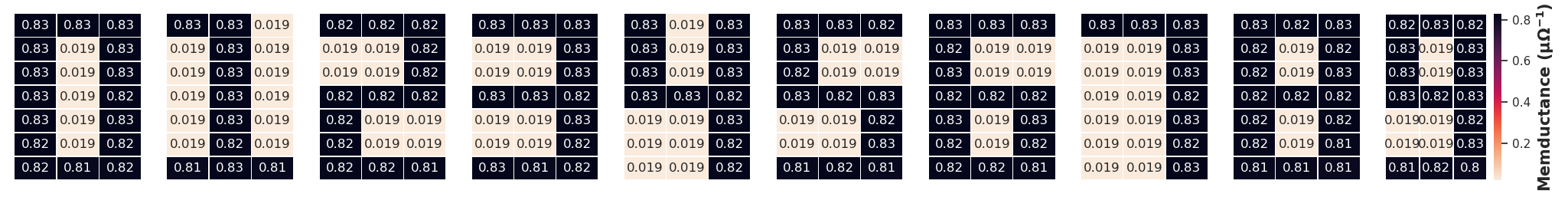}
    \caption{In order to scale the SNN in terms of input and output neurons, ten binary digits 0 to 9 are taken to train. The final weights after training consist of digits 0 to 9 of size $7 \times 3$.}
    \label{fig:ten}
\end{figure}

The memristive SNN is scalable due to its modular architecture and can adapt to various pattern sizes and numbers. We illustrate this scalability by training the SNN with ten binary images (0 to 9) of size \(7 \times 3\) with $21$ pre- and 10 post-neurons. The heat map of the synaptic weights after training, presented in Fig.~\ref{fig:ten}, agrees with the patterns in the scaled memristive SNN. During testing, all patterns are recognized correctly.

\subsection{Simulation for Classification}
The inputs of the BCW and IRIS data sets are normalized first by Min-Max scaling, then pre-processed by applying Gaussian receptive fields~\cite{sboev2020solving} to increase the number of features three times. Originally, IRIS and BCW have $4$ and $30$ features, but they have increased to $12$ and $90$ after pre-processing, respectively. Therefore, the number of pre-spikes for IRIS and BCW is $12$ and $90$, respectively. The number of post-neurons or output neurons is three and two in the memristive SNN for IRIS and BCW, respectively. After training, we got the classification accuracy of $99.11\%$ and $97.9\%$ for the IRIS and BCW data sets, respectively. The comparison of F1-score and classification accuracies for IRIS and BCW with previous SNNs solving classification problems is presented in \autoref{tab:f1score}. The comparison shows that our memristive SNN gives better results. It is worth noting that in this work, the number of features increased by three times after pre-processing, but in work~\cite{sboev2021mathematics}, the same increases by twenty times.

Compared with Jiang \etal~\cite{jiang2022memristor}, the proposed design exhibits lower circuit complexity. Jiang \etal employ a multi-synaptic architecture with multi-delay pulse transmission and hidden layers, whereas the proposed architecture utilizes a single-synapse 1T1M structure without delay components or hidden layers, simplifying circuit and routing complexity. Each LIF neuron in~\cite{jiang2022memristor} requires a 555-timer circuit with three capacitors, four resistors, and two voltage-controlled switches, while the proposed neuron circuit employs one capacitor, one resistor, one operational amplifier, one reset voltage source, and two voltage-controlled switches. Despite lower hardware complexity, the proposed architecture supports in-situ supervised learning and achieves $99.11\%$ classification accuracy on the IRIS dataset versus $97.33\%$ in~\cite{jiang2022memristor}.

\begin{table*}[]
\caption{Comparison of F1-score and classification accuracy for different datasets with previous works. }
\label{tab:f1score}
\resizebox{\linewidth}{!}{%
\begin{tabular}{|c|cc|cc|c|c|c|}
\hline
\multirow{2}{*}{Papers} &
  \multicolumn{2}{c|}{F1-score} &
  \multicolumn{2}{c|}{Accuracy} &
  \multirow{2}{*}{\begin{tabular}[c]{@{}c@{}}Memristive\\ Architecture\end{tabular}} &
  \multirow{2}{*}{\begin{tabular}[c]{@{}c@{}}Circuit level\\ design\end{tabular}} &
  \multirow{2}{*}{\begin{tabular}[c]{@{}c@{}}Simulated on\\ Circuit\end{tabular}} \\ \cline{2-5}
                            & \multicolumn{1}{c|}{IRIS}      & BCW        & \multicolumn{1}{c|}{IRIS}      & BCW      &     &     &     \\ \hline
\cite{sboev2020solving}     & \multicolumn{1}{c|}{$99\%$}    & $90\%$     & \multicolumn{1}{c|}{---}       & ---      & No  & No  & No  \\ \hline
\cite{sboev2021mathematics} & \multicolumn{1}{c|}{$97\%$}    & $90\%$     & \multicolumn{1}{c|}{---}       & ---      & Yes & No  & No  \\ \hline
\multicolumn{1}{|c|}{\cite{jiang2022memristor}} &
  \multicolumn{1}{c|}{$97.326\%$} &
  \multicolumn{1}{c|}{---} &
  \multicolumn{1}{c|}{$97.33\%$} &
  --- &
  \multicolumn{1}{c|}{Yes} &
  \multicolumn{1}{c|}{Yes} &
  No \\ \hline
{\bf This work}             & \multicolumn{1}{c|}{$98.68\%$} & $98.617\%$ & \multicolumn{1}{l|}{$99.11\%$} & $97.9\%$ & Yes & Yes & Yes \\ \hline
\end{tabular}
}
\end{table*}


\subsection{Hardware Non-Idealities and Variability Analysis}


Physical memristive crossbars are vulnerable to fabrication faults and device non-idealities, motivating the analysis of their impact on memristive SNN performance. Accordingly, the proposed architecture is evaluated under the following hardware non-idealities and device variations:


\subsubsection{Stuck-at-a-conductance state}
The physical memristor may be stuck at a conductance state and does not change its state even on the application of voltages greater than its threshold across it. 
We performed simulations for the IRIS and BCW datasets by choosing random $5\%,~10\%$, and $20\%$ of memristors that are stuck at random conducting states. With these faulty memristors, the training of memristive SNNs is performed with the \autoref{algo:weightUpdate} and results in the accuracies of $96.44\%$, $97.33\%$, and $92.88\%$ for IRIS as well as $96.50\%$, $96.50\%$, and $96.50\%$ for BCW data sets, respectively.  The results demonstrate that the proposed architecture is robust against these faults. It is to be noted that the performance for the BCW data set does not change much because of its memristive SNN size compared to that of IRIS. 

\subsubsection{Variations in boundary resistances $R_{on}$ and $R_{off}$}
The random variations in the memristor's boundary resistance limits are added. The variation is defined by relative standard dispersion, i.e., the ratio between variation and mean value of the parameter\cite{zhao_TCAS_2020}. The memristive SNN for IRIS data are trained by adding $5\%$, $10\%$, $20\%$, $25\%$, and $30\%$ variations in the limits of both $R_{on}$ and $R_{off}$ of all memristors~\cite{Zhang_tcas_2017} of the MC and got $98.22\%$, $97.33\%$, $94.66\%$, $96\%$, and $81.34\%$ accuracies respectively. The accuracies are evidence of the resilience of memristive SNN against these variations.

\subsubsection{Variations in memristor threshold voltages}
The memristor's threshold is very crucial for synaptic plasticity because the memristor's conductance is only changed if it experiences voltages greater than the threshold. The threshold parameters suffer from device-to-device variation. We analyzed this variation using the memristive SNN with the IRIS dataset. In order to consider this, we added $5\%$ and $10\%$ variations in positive and negative thresholds of all memristors of MC. After training, it gives $97.33\%$ and $76.89\%$ accuracies, respectively. The results illustrate that the proposed model is sensitive to variations in the memristor's thresholds.

\subsubsection{Cycle-to-cycle variation}

Physical memristors exhibit cycle-to-cycle (C2C) variation, i.e., stochastic conductance fluctuations under identical programming conditions~\cite{eslami2026efficient, BISCHOFF2022108321}. To capture this effect, LTspice simulations on the IRIS dataset were performed by introducing a Gaussian stochastic term into the memristor state dynamics (\autoref{dynamic}), yielding
\begin{equation*}
\mu_{v}^{\text{effective}}=\mu_v(1+\eta_k),
\end{equation*}
where $\eta_k\sim\mathcal{N}(0,\sigma_{c2c}^2)$ is a zero-mean Gaussian random variable with standard deviation $\sigma_{c2c}$. The model was evaluated for $\sigma_{c2c}\in\{0.1,0.2\}$, with $\eta_k$ kept constant during the conductance update for sample $k$ to ensure consistent C2C behavior. The proposed model achieved $97.33\%$ accuracy for both variability levels, demonstrating strong resilience against C2C variation.

\subsubsection{Line resistance}

In practical MC implementations, interconnect wires introduce finite line resistance, causing IR-drop-induced voltage degradation along rows and columns~\cite{Liuiccad2014}. To evaluate this effect, resistive segments were inserted along both row and column interconnects of the MC in LTspice simulations on the IRIS dataset. Two cases were considered: (i) $5\,\Omega$ and (ii) $10\,\Omega$ per segment. The obtained classification accuracies were $98.22\%$ and $97.33\%$, respectively, demonstrating stable performance and robustness against line-resistance induced IR-drop effects.

\subsubsection{Parasitic capacitance Analysis}

In practical MC implementations, finite interconnect resistance and parasitic capacitance introduce IR drop and RC delay along rows and columns~\cite{ThomasMWSCAS2021, mi12070791}, potentially distorting spike and programming signals. To evaluate robustness against these parasitics, distributed resistive and capacitive elements were incorporated along both row and column lines of the MC. Two resistance values ($R=5\,\Omega$ and $10\,\Omega$ per segment) and two parasitic capacitance values ($C=5\,\mathrm{fF}$ and $10\,\mathrm{fF}$) were considered in LTspice simulations using the IRIS dataset. The proposed memristive SNN achieved $97.33\%$ classification accuracy for all four $R$-$C$ combinations, demonstrating resilience against practical interconnect parasitic effects.

\subsection{Discussion}
The memristive SNNs circuits and in-situ algorithm together are a complete system that works without any separate control unit, like a micro-controller or FPGA, etc. The model is modular, scalable, and shows great robustness against device non-idealities and noisy patterns.
From \autoref{tab:comparison}, it is efficient in hardware complexity, which reduces area and energy consumption. For classification, the memristive SNNs with IRIS and BCW datasets took only $10$ and $5$ epochs to train them, respectively, with comparable accuracies in \autoref{tab:f1score}.

In this work, the threshold voltage of the NMOS transistor in 1T1M synapses is $0.5V$. The gate voltage is set to either +5V or -5V to switch the NMOS ON or OFF, respectively, resulting in the synapse being either active or inactive. At the drain terminal, we apply either a spike (maximum 1.1V) or 1.4V or -2.6V. In these conditions, the NMOS synapse operates either in the cutoff or triode modes, but not in saturation. The total consumed power at a 1T1M synapse is the sum of the powers that NMOS and memristor consume individually. The power consumed by the NMOS transistor is the product of the drain current and the drain-source voltage when it is not in cut-off mode~\cite{weste2015cmos}. When the transistor is in the cut-off mode \cite{wang2021mosfets}, no current passes through the NMOS transistor and memristor of the synapse, resulting in zero power consumption. Therefore, power is consumed only when the synapse is active, and the NMOS transistor is in the triode mode.


System-level analysis of the proposed memristive SNN is performed in terms of energy, latency, and hardware complexity. During training, each synapse consumes an average energy of $21.878\,\mu\mathrm{J}$ per sample, while the total average energy consumption of the $12\times3$ MC is $0.397\,\mathrm{mJ}$. The lower-than-linear scaling arises from the event-driven nature of the SNN and the WTA mechanism, which restricts synaptic updates to the winning neuron, thereby reducing simultaneous switching activity. The processing latency per input sample, including spike propagation and synaptic programming, is $1\,\mathrm{ms}$. Hardware complexity is discussed in terms of synapse count and associated peripheral circuits, as summarized in \autoref{tab:comparison}.

\begin{figure}[h]
    \centering
    \includegraphics[width=0.8\linewidth]{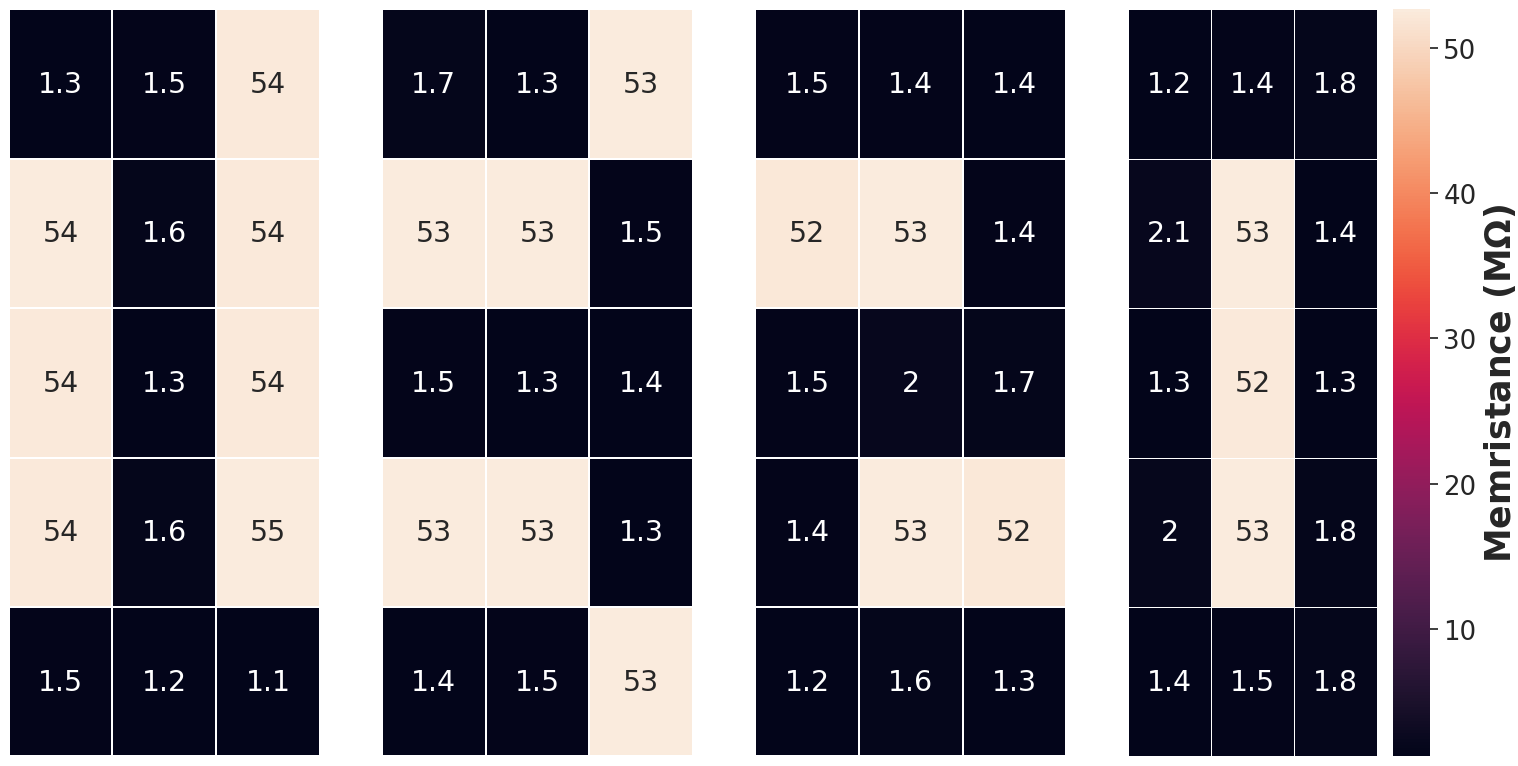}
    \caption{Heat map of synaptic weights illustrating the association between output neurons and input pattern types after unsupervised training of the memristive SNN. Each neuron exhibits selective activation toward a distinct pattern despite the absence of supervision.}
    \label{fig:unsupervised}
\end{figure}

During training of the model for a pattern recognition task, it is observed that in the absence of bias current ($I_b = 0$) and with memristors randomly initialized near their low-resistance states, each output neuron becomes associated with a specific input pattern type, as illustrated in Fig.~\ref{fig:unsupervised}. This neuron-pattern association emerges in a random manner. The heat map shown in Fig.~\ref{fig:unsupervised} indicates that $neuron_{0}$, $neuron_{1}$, $neuron_{2}$, and $neuron_{3}$ predominantly respond to $pattern_1$, $pattern_3$, $pattern_2$, and $pattern_0$, respectively. Notably, this process does not require label information during training, thereby demonstrating an unsupervised learning behavior.

Unsupervised behavior was also evaluated on the IRIS dataset. It was observed that the third neuron predominantly spikes for samples of the first class, the second neuron mainly responds to samples of the second class and a few samples of the third class, while the first neuron primarily spikes for the third class. This behavior demonstrates spontaneous neuron specialization and competitive self-organization without externally imposed target supervision.

The 1T1M synapses significantly suppress sneak-path currents typically present in passive MC arrays~\cite{yakopcic_IJCNN_2011}. The access transistor electrically isolates unselected memristors during synaptic update operations, preventing unintended current paths and improving programming reliability. During spike transmission, i.e., prior to synaptic weight updates, synaptic currents propagate through the MC, enabling each neuron to receive the weighted sum of currents from all connected synapses.

In order to examine the trade-off between accuracy and hardware complexity, a reduced feature-expansion pre-processing is evaluated~\cite{sboev2020solving}. While 3× feature expansion yields 99.11\% accuracy on the Iris Dataset, reducing the expansion to 2× lowers the accuracy to 89.33\%. This indicates that feature expansion improves classification performance but increases the number of synaptic elements in the memristive crossbar, highlighting the inherent trade-off between accuracy and hardware complexity.

To examine the trade-off between accuracy and hardware complexity, reduced feature expansion was evaluated~\cite{sboev2020solving}. While 3$\times$ feature expansion achieves $99.11\%$ accuracy on the IRIS dataset, reducing the expansion to 2$\times$ lowers the accuracy to $89.33\%$. This demonstrates that feature expansion improves classification performance at the cost of increased synaptic elements in the memristive crossbar, highlighting the trade-off between accuracy and hardware complexity.

Future work will focus on deeper memristive SNN architectures and explore their deployment for larger datasets in edge-intelligent devices.

\section{Conclusion}

\label{conclusion}
This work presents a hardware-efficient memristive SNN architecture with an STDP-based supervised in-situ training algorithm. The architecture efficiently incorporates lateral inhibition and refractory mechanisms, while enabling concurrent synaptic updates for the winner neuron. The proposed modular architecture is robust against stuck-at faults, device variations, noisy input patterns, cycle-to-cycle variability, line resistance, and parasitic capacitance. The memristive SNN achieves classification accuracies of $99.11\%$ and $97.9\%$ on the IRIS and Breast Cancer Wisconsin datasets, respectively. Under $20\%$ randomly stuck memristors, the model maintains accuracies of $92.88\%$ for IRIS and $96.5\%$ for BCW. Moreover, the architecture achieves $81.34\%$ accuracy under $30\%$ variation in memristor limit resistances and maintains at least $97.33\%$ accuracy on the IRIS dataset under practical interconnect and device non-idealities, demonstrating strong robustness and stability.
\ifCLASSOPTIONcaptionsoff
  \newpage
\fi



%
\bibliographystyle{IEEEtran}
\bibliography{main.bib}



\begin{IEEEbiography}[{\includegraphics[width=1in,height=1.25in,clip,keepaspectratio]{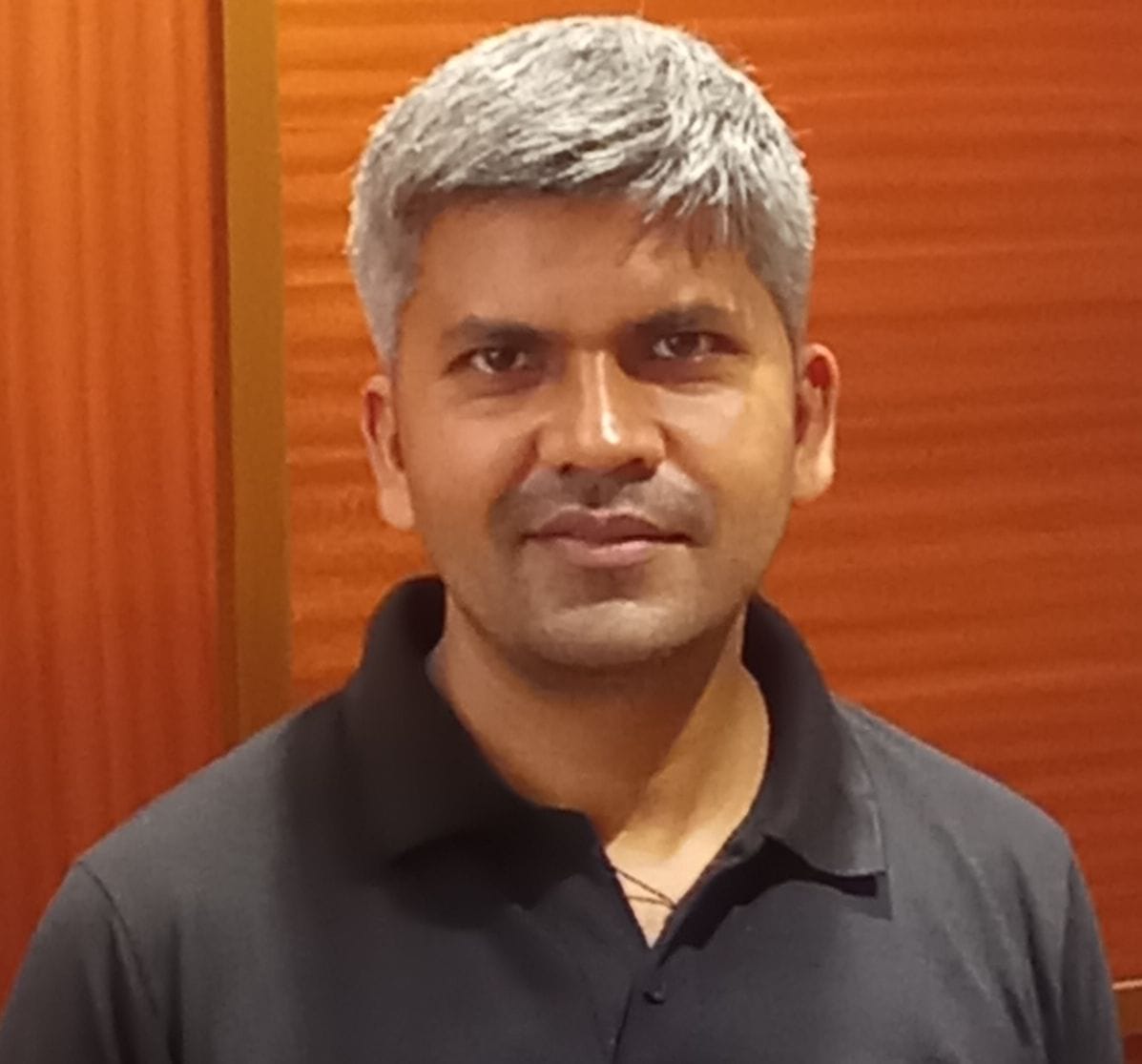}}]{Santlal Prajapati}
is Ph.D. scholar of Indian Statistical Institute, Kolkata India. He received his master degree in Computer Science and Engineering from Jadavpur University, Kolkata, India in , and B.Tech degree from Academy of Technology, Hooghly, India in 2012. He received Gate and UGC Net scholarships in 2012 and 2017 respectively. He served as a faculty from November 2015 to July 2018 in Indian Statistical Institute, Giridih. Currently, He is assistant professor of Sister Nivedita University, Kolkata. His research interests include neuromorphic computing, memristor-based circuits, in-memory computing, Non-volatile memory, and artificial intelligence.

\end{IEEEbiography}

\begin{IEEEbiography}[{\includegraphics[height=1.0in,clip,keepaspectratio]{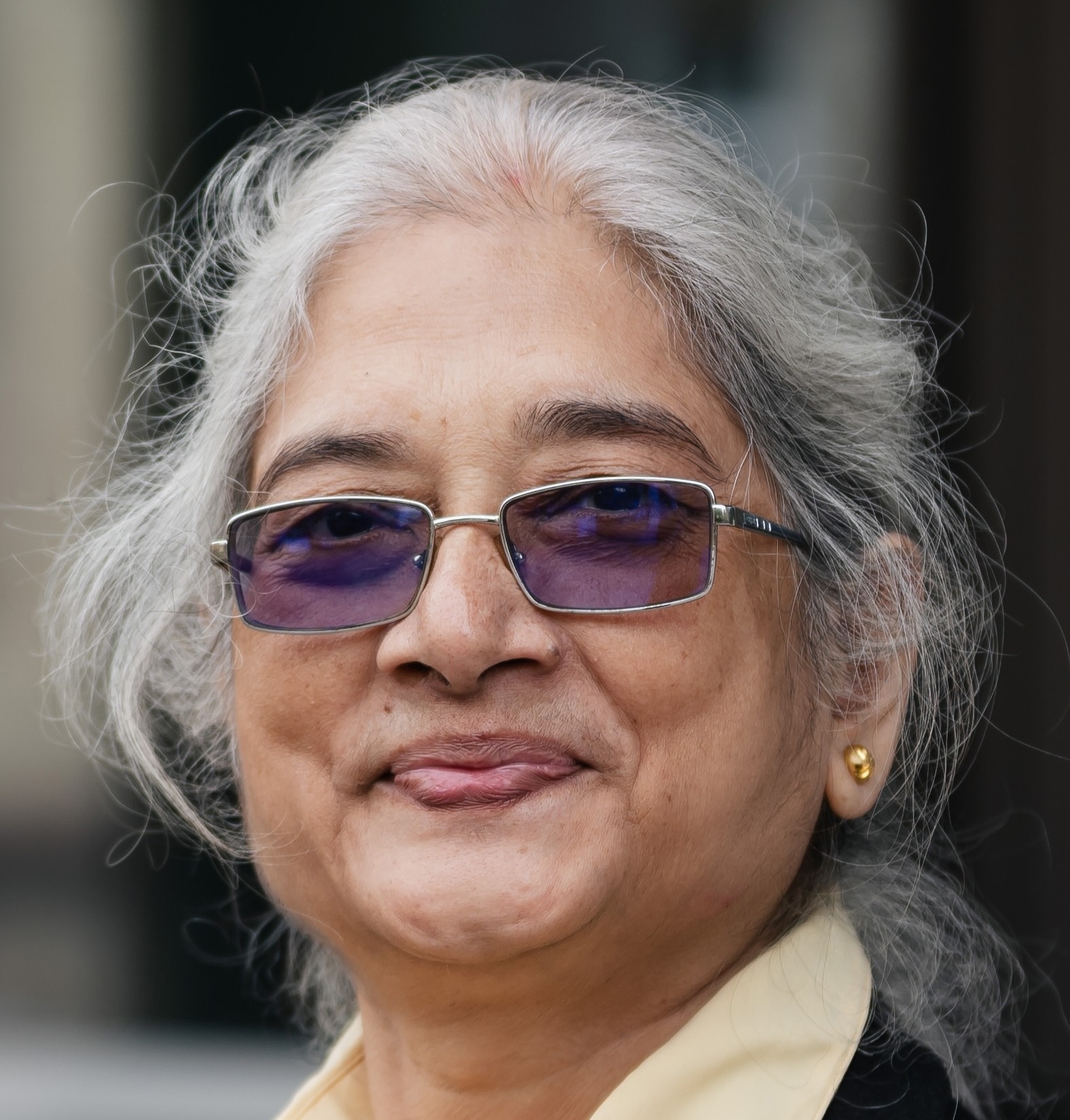}}]{Susmita Sur-Kolay}  was a
Professor at Indian Statistical Institute and recently a Fulbright Academic Excellence Fellow in the CSE Department of the University of California, San Diego. Her research
interests include design automation for nanoelectronic and quantum circuits, hardware security and
in-memory neuromorphic computation. She received
her Ph.D. degree in Computer Science and Engineering from Jadavpur University, and from 1980 to
1985, she was at Massachusetts Institute of Technology. She is a Fellow of
Indian National Science Academy and Indian National Academy of Engineering, a recipient of the President of India Gold Medal and Distinguished
Alumnus Award of IIT Kharagpur, IEEE Women in Engineering Inspiring Member of the Year Award and Women in Technology Leadership
Award of the VLSI Society of India.
\end{IEEEbiography}

\begin{IEEEbiography}[{\includegraphics[width=1in,height=1.25in,clip,keepaspectratio]{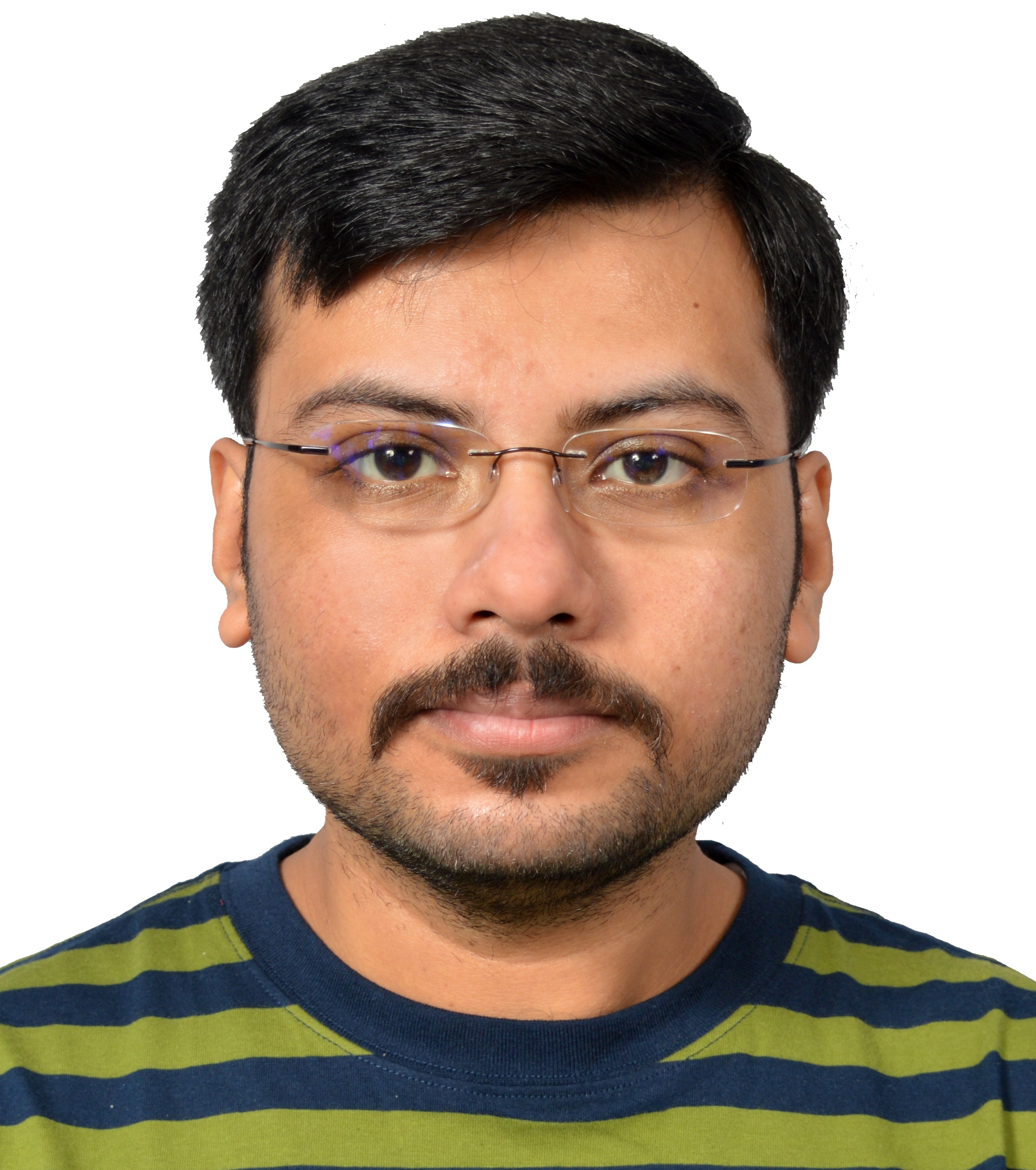}}]{Soumyadeep Dutta}

received the B.Tech. degree in Electronics and Communication Engineering from the National Institute of Technology, Durgapur, India, in 2009, and the Ph.D. degree in Applied Physics from the Indian Institute of Science, Bengaluru, India, in 2018, specializing in memory materials and resistive switching in perovskite oxide thin films. He held Research Associate positions at the Indian Institute of Science and the Indian Institute of Technology Madras, working on defect physics and electronic transport in resistive-switching materials. He was an Experienced Associate (Data Scientist) with PricewaterhouseCoopers (PwC), Kolkata, India, from 2021 to 2023, and later served as an independent AI consultant. He is currently a Data Science Manager with Balasore Alloys Limited, Kolkata, India, working on multi-agent AI architectures and time-series analytics for industrial process control. His research interests include condensed matter physics, computer vision, natural language processing, and neuromorphic computing.

\end{IEEEbiography}










\end{document}